\documentclass[epsfig,preprint,showpacs,12pt]{revtex4}
\input{epsf.tex}
\usepackage{amssymb,epsfig}
\newcommand{\One}{1\kern-4.5pt1}

\begin{document}
\renewcommand{\topfraction}{.95}
\renewcommand{\textfraction}{.1}

\title{Zero-temperature phase transitions in dilute bosonic superfluids on a lattice}

\author{A. S. Alexandrov and I. O. Thomas}
\affiliation{Department of Physics, Loughborough University, Loughborough LE11 3TU, United Kingdom}

\begin{abstract}
Kinetic energy driven phase transitions in Bose superfluids occur at low values of the repulsion when the values of the next-to-nearest and next-to-next-to-nearest hopping term attain certain critical values, resulting in alterations in the wave vector of the condensate.  We map out the space of possible phases allowed by particular forms of the single-particle energy dispersion in the superfluid state, noting the appearance of a new phase, and examine in more detail the effects of additional repulsive terms on the form of the condensate wavefunction.  We also examine the effect of these additional hopping terms on the formation of inhomogeneities in the condensate.
\end{abstract}
\maketitle

\section{Introduction}\label{sec:Intro}  

Ever since the phase transition (QPT) from a superfluid liquid to a Mott insulator was observed in atomic gases trapped in an optical lattice \cite{Grenier:2002}, there has been considerable experimental and theoretical activity geared towards the exploration of the physics of this transition as described by the Bose-Hubbard model \cite{Jaksch:1998}.  This phenomenon has also long been of interest in the study of phase transitions of charged Bose liquids as observed in the context of granular superconductors and Josephson junction arrays \cite{Fisher:1989}, in addition to preformed real-space electron pairs such as bipolarons \cite{Alexandrov:1994}.

The superfluid-Mott transition occurs only at fixed densities in a homogeneous system when the number of bosons is commensurate with the number of lattice sites.  At intermediate densities a transition may occur to a density wave superfluid \cite{Alexandrov:1981,Kubo:1983} (similar to the putative supersolid phase of $^4$He \cite{Chan:2008}) which has a tendency towards phase separation \cite{Aubrey:2007,Batrouni:2000,Moskvin:2004}.  The transitions require that both the repulsive interaction and the number of bosons is sufficiently large: the critical value of this repulsion for commensurate density has been determined numerically for two-dimensional square lattices \cite{Krauth:1991,Krauth:1992,Cappello:2007,Capogrosso-Sansone:2008,Cappello:2008}.  

Alexandrov and Thomas \cite{Alexandrov:2008} have shown that the inclusion of modifications to the kinetic energy term of the tight-binding Gross-Pitaevskii \cite{Gross:1961,Pitaevskii:1961} equation (a mean field description of the Bose-Hubbard Model \cite{Amico:2000}), along with additional longer-range repulsive terms gives rise to condensate stripes or checkerboards (as anticipated for d-wave superconductors in \cite{Alexandrov:2003}) with wave-vectors that are not necessarily commensurate with the lattice period.  Unusually, it is the values of the additional hopping parameters that drive these `kinetic' phase transitions between the homogeneous phase and the striped/checkerboard phases, and not the repulsive potentials -- though the repulsion does play a role regarding whether stripes or checkerboards are formed.  In the present paper, we expand on those results, examining these phenomena in more detail. 

At first glance, it might not seem that one could expect a phase transition in the ground state of a near-ideal Bose gas whose inter-particle repulsion is small compared to their kinetic energy.  In particular, this absence of any significant effects is suggested by the rather strong constraints placed on the form of the bosonic ground state wave function by the requirement that it posses no nodes \cite{Landau:1977,Courant:1989}, which implies that the values of the hopping terms for the tight-binding form of the s-wave ground state must be such that they approximate a kinetic dispersion with a single minimum located at the zero wave-vector.  There are a number of subtleties, however, which appear to indicate that this is not all that can be said on the matter.

Firstly, there exist some cases (such as that of bipolarons) where the condensation occurs in the p-band (or the d-band in the case of single polarons), and thus one's choice of hopping parameters is no longer limited by the no-node restriction, since this is not the ground state of the system as a whole.  Another potential example of this is the incommensurate, metastable superfluidity of atoms in the first excited band of a double-well optical lattice that has been predicted by Stojanovi\'c {\em et al.} \cite{Stojanovic:2008}, since it has been shown that the lifetimes of excited atoms in p-wave orbitals in optical lattices may be extended by an unexpectedly large amount \cite{Isacsson:2005,Muller:2007}.  

Secondly, the assumptions of this no-go theorem \cite{Courant:1989} are violated in the presence of significant many-body effects.  This can be explicit, as in the case of the non-linear potential in the GPE (which renders the Hamiltonian non-self-adjoint), or implicit and therefore hidden in the form of the hopping terms such that they no longer approximate the continuum Laplacian operator of the kinetic term.   This last can occur, for example, as a result of interactions with other systems that produces long-range dispersive effects such as in Keverekidis {\em et al.}'s \cite{Kevrekidis:2001} study of the dynamics of discrete breathers in a coupled linear and non-linear system, or such as Feshbach resonance-type effects that result in the {\em p-wave} state becoming the system ground state \cite{Kuklov:2006,Liu:2006}.  Similarly, Larson and Martikainen have shown that the interaction between internal and external degrees of freedom in an ultra-cold gas of atoms can give rise to a transition to a ground state with nodes as the detuning of the atom field is adjusted \cite{Larson:2008a}; when written in terms of an effective Bose-Hubbard model, this new ground state corresponds to a negative value of the nearest-neighbour hopping  parameter \cite{Larson:2008b} -- that is, the dispersion has a p-wave form, characterised by an `antiferromagnetic' ordering of the condensate wave function.

Given the above considerations, it is worth exploring the behaviour of bosonic superfluids on a lattice in the GPE limit for all relevant values of the additional hopping parameters, despite possible difficulties in the straightforward experimental replication of these values on certain optical lattices \cite{Scarola:2008}. Similar analytic and numerical analyses have been performed in the case of the full Bose Hubbard model \cite{Buonsante:2004, Chen:2008}, though not for precisely identical parameters.  In what follows, we outline our formalism and numerical method (\S\ref{sec:procedure}), expand on the notion of kinetic energy driven phase transitions (\S\ref{sec:KPTout}) first described in \cite{Alexandrov:2008} and noting the appearance of a new phase not described earlier, before discussing our numerical results in the context of this theory (\S\ref{sec:swallowtail} and \S\ref{sec:results}).

\section{The tight-binding Gross-Pitaevskii equation}\label{sec:procedure}

We begin with a system whose  ground state  is described by a
Gross-Pitaevskii-type (GP) equation \cite{Gross:1961,Pitaevskii:1961}
including lattice, $V({\bf r})$, and  interaction, $U({\bf r})$,
potentials,
\begin{equation}
\left[-{\hbar^2 \nabla^2 \over{2m}}+V({\bf r}) -\mu +  \int d{\bf
r'} U({\bf r} -{\bf r'})|\psi ({\bf r'})|^2\right] \psi({\bf r}) =0,
\label{eqn:gp}
\end{equation}
\noindent where the condensate wave-functions $ \psi({\bf r})$ play the role of
the order parameter  $m$ is the boson mass, and $\mu$
is the chemical potential.  $\psi({\bf r})$  is normalised through$\int d{\bf r} |\psi({\bf r})|^2 =N_b$, where $N_b$ is the number of bosons determined by the value of $\mu$.

To solve equation (\ref{eqn:gp}) variationally on a 
lattice with a large number $N$ of sites, we make use of a complete set of
orthogonal Wannier (site) functions $w({\bf r})$. Writing the 
order parameter in terms of these as  $ \psi({\bf r})= \sum_{\bf m} \phi _{\bf m} w
({\bf r-m})$, we can reduce (\ref{eqn:gp}) to a discrete set of equations for
the amplitudes at each lattice site, $\phi _{\bf m}$,
\begin{equation}
-\sum_{\bf m} [t({\bf m-m'})+\mu \delta_{\bf m,m'}] \phi _{\bf m} +
\sum_{\bf m, n,n'}U_{\bf mn}^{\bf m'n'} \phi^* _{\bf n'}\phi _{\bf
n}\phi _{\bf m} =0. \label{eqn:gpsite}
\end{equation}
Here $t({\bf m})=\int d{\bf r} w^*({\bf r})[-\hbar^2 \nabla^2/2m
+V({\bf r})] w({\bf r-m})$
 is the hopping integral, and  $U_{\bf m n}^{\bf m'n'}=\int \int d{\bf r}d{\bf r'}U({\bf
r-r'}) w^*({\bf r-m'})w^*({\bf r'-n'})w({\bf r'-n})w({\bf r-m})$
 is the matrix element of the interaction potential.

In the liquid regime far away from the superfluid-Mott transition, the hopping integrals dominate over the
interaction matrix elements, so that we need only keep
the density-density interactions, $U_{\bf mn}^{\bf m'n'}\approx
U_{\bf mn}^{\bf m n}\delta_{\bf m,m'}\delta_{\bf n,n'}$. Solutions to the GP equation (\ref{eqn:gpsite}) are then equivalent to the minima
of the energy functional,
\begin{equation}
E(\phi _{\bf m})=-\sum_{\bf m,n} \left[t({\bf m-n})+\mu \delta_{\bf
m,n}
 - {1\over{2}}U_{\bf m
n}^{\bf m n} \phi^* _{\bf m} \phi _{\bf n}\right]\phi^* _{\bf n}\phi
_{\bf m}. \label{eqn:functional}
\end{equation}

Rescaling the order parameter as $\phi _{\bf m}=n^{1/2}
f_{\bf m}$ and the interaction as $U_{\bf mn}^{\bf m n}= u_{\bf
nm}/n$ yields a universal functional, $\tilde{E}(f_{\bf m})=E(\phi
_{\bf m})/n$ {\em that is independent of} $n=N_b/N$, the number of bosons per
lattice valley: 

\begin{equation}
\tilde{E}(f_{\bf m})=-\sum_{\bf m,n} \left[t({\bf m-n})+\mu
\delta_{\bf m,n}
 - {1\over{2}}u_{\bf m
n} f^* _{\bf m} f _{\bf n}\right]f^* _{\bf n}f _{\bf m}.
\label{eqn:functional2}
\end{equation}
Importantly, this shows that solutions for different particle densities can be mapped on to each other by a simple rescaling of the interaction.

\begin{figure}[htb]
\vspace{0.5cm}
\begin{center}
\epsfig{file=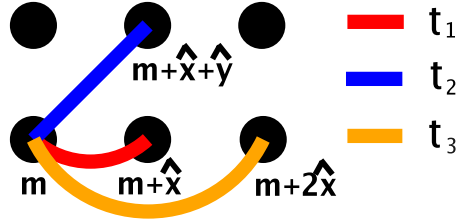,height=3.0cm}
\end{center}
\vspace{-.5cm}
\caption{\small Diagram showing the different hopping parameters from a given site in the forward right direction on a 2D lattice.}
\label{fig:hopping}
\end{figure}

In what follows we examine the effects of the variation of the values of $t_2$ and $t_3$, the next-to-nearest neighbour and the next-to-next-to-nearest neighbour parameters respectively (Fig. (\ref{fig:hopping})), for various small values of the onsite ($u_{00}$), nearest neighbour ($u_{01}$) and next-to-nearest neighbour ($u_{02}$) repulsive potentials.  

\begin{figure}[htb]
\begin{center}
a)
\epsfig{file=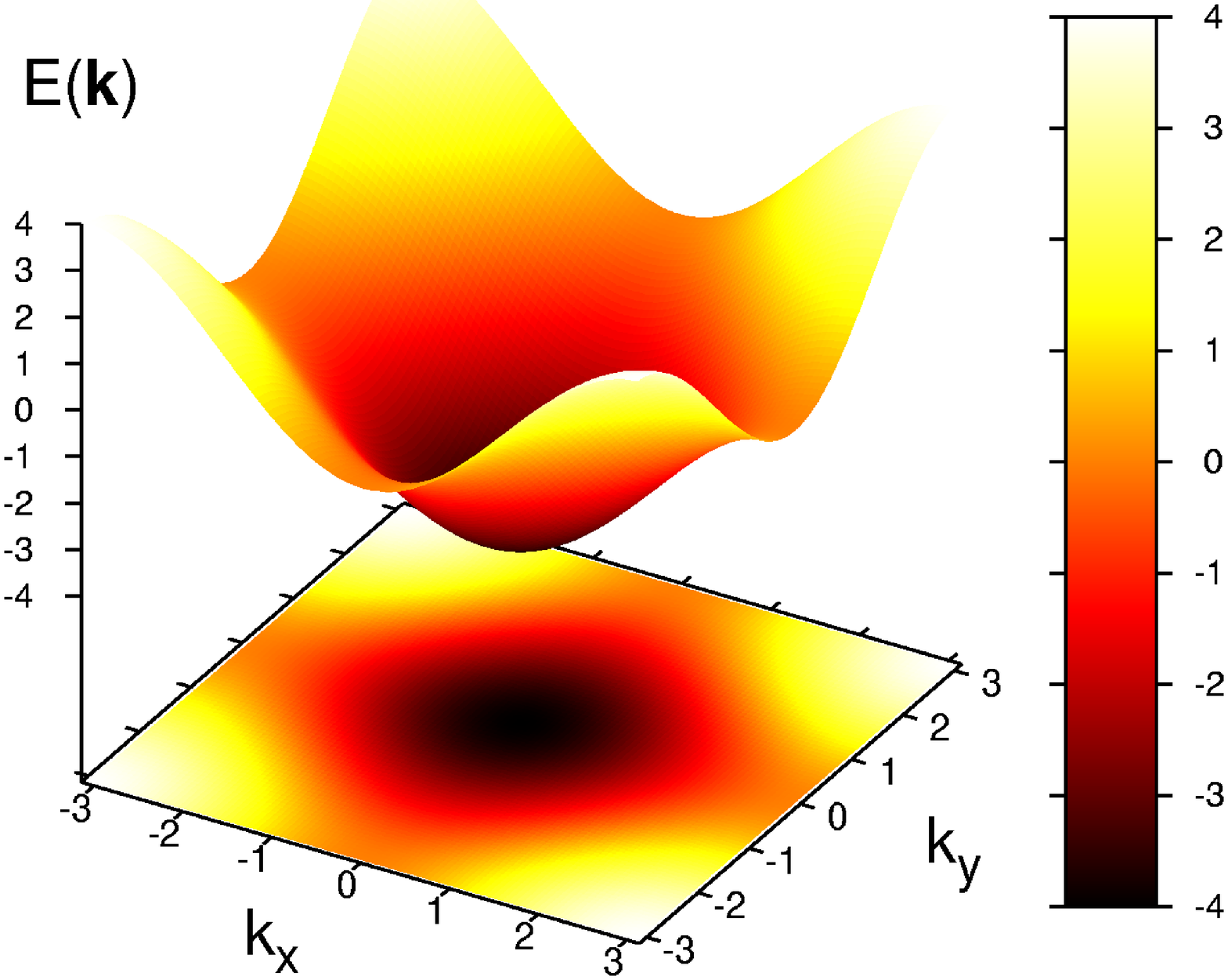,height=5.4cm}
\hspace{.5cm}
b)
\epsfig{file=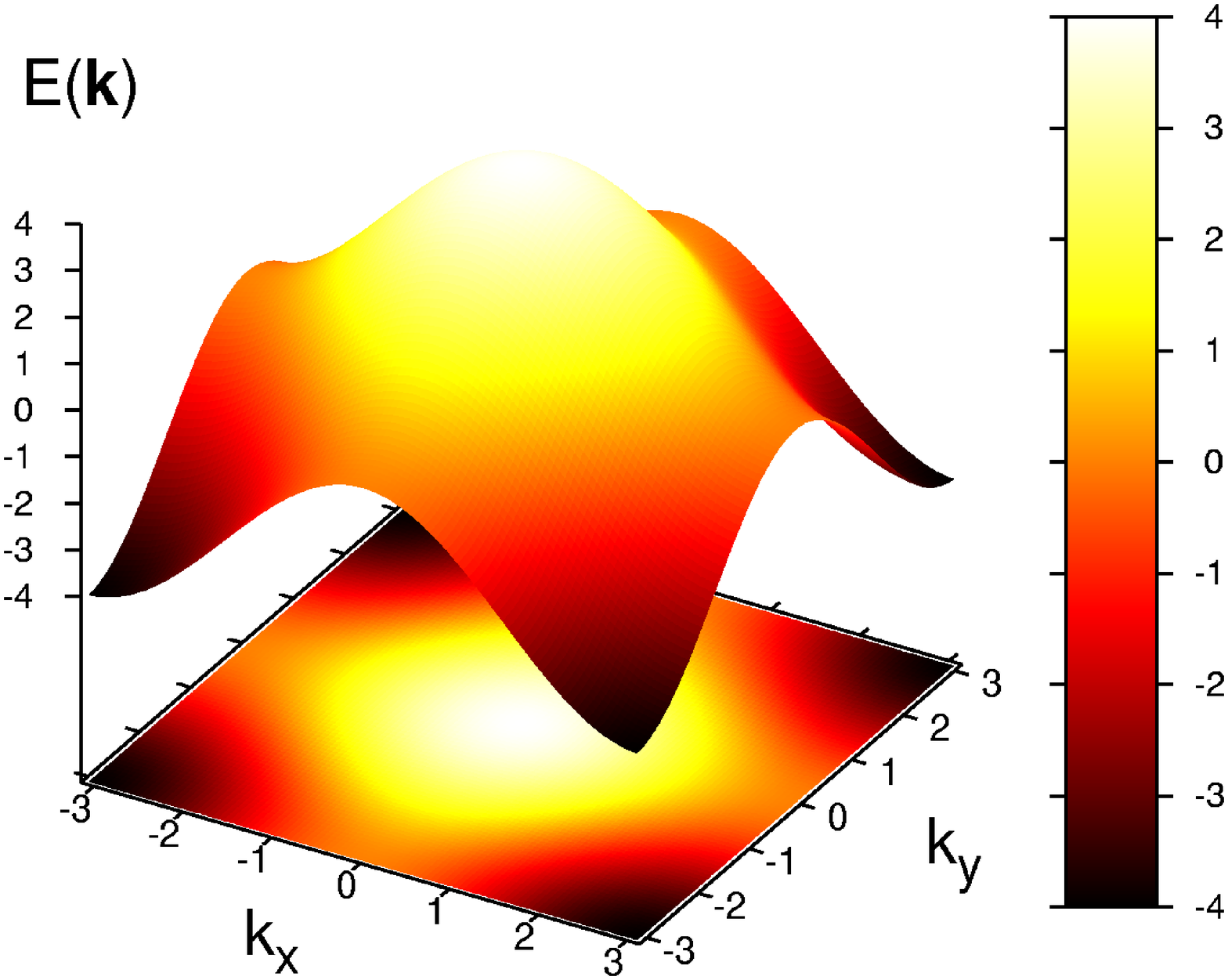,height=5.4cm}
\end{center}
\vspace{-.5cm}
\caption{\small Dispersion relations for a) conventional ($t_1=1$) and b) unconventional ($t_1=-1$) cases with $t_2=0$ and $t_3=0$.}
\label{fig:Denergy1}
\end{figure}

Numerical results were obtained through the following procedure:  we generated 
random starting values of the amplitudes $f_{\bf m}$ (where $f_{\bf m}\in \Re$ because there is no current) on a lattice with $N$ sites and periodic boundary conditions. The functional (\ref{eqn:functional2}) was minimised for a given value of $\mu$ using the NAG conjugate gradient optimisation routine E04DGF.  When calculated as part of this procedure, the  kinetic portion of the functional was calculated in momentum space following a Fourier transformation of the amplitudes at each lattice site.  This was then inverse Fourier transformed back into Wannier space where the non-linear potential portion of the functional was then applied.  Note that the algorithm tends to fail in the absence of a small on-site repulsion.  We fine-tuned the value of $\mu$ using the Van Wijngaarden-Dekker-Brent algorithm \cite{Press:1992} so that the constraint $\sum_{\bf m}|f_{\bf m}|^2=N$ was imposed; typically $\mu$ is determined to an accuracy of $10^{-7}-10^{-9}$, corresponding to a relative error of  order $10^{-6}-10^{-7}$ in the enforcement of the constraint, which is  sufficient for our purposes.  Unless otherwise stated, a square 2D lattice of $N=50^2$ has been used for the plots of the condensate function; simulation on a variety of lattice sizes seems to indicate that this is most likely large enough to avoid significant finite size effects.  However, $25\times25$ lattices seem to give good  qualitative indications of the phase at a given point, which should be largely independent of the lattice size.  Due to the relatively quick speed with which the minimisation algorithm runs for lattices of this size, the results in Figure \ref{fig:Chkr1} pertaining to the scanning of the $t_2$ and $t_3$ parameter spaces were generated on these.  Some cross-checks for certain parameter sets were performed on $50\times50$ lattices; when in doubt, these were used to determine the phase of the condensate.

It is worth noting that one sometimes finds in the ordered phases that a pair of parallel `kinks' (or a quartet, if they are ordered diagonally) has formed in the condensate, and that these are often slightly more stable than the ordered state alone.  We suspect that these are artefacts of the periodic boundary conditions and the finite lattice size, perhaps made frequent by our use of a random start instead of a trial wave-function; firstly because at small repulsions, the `kinks' vanish -- or become much less common -- when the lattice size is set to $50\times50$, and secondly because they always appear in pairs or quartets so as to satisfy the boundary conditions.  We therefore neglect them; even if they are not artefacts the presence of `kinks' in the order of the condensate are not as important to us as the nature of that order, which is the focus of our study, and they are qualitatively different from the meandering domain walls discussed in \S\ref{sec:swallowtail}.

In addition, on the $25\times25$ lattices one finds that in the commensurate phases there is a dislocation; this is an artefact both of the periodic boundary conditions and there being an odd number of lattice sites in each direction, since this entails that the wavefunction at one end is out of phase with the wavefunction at the other end.  This is not an obstacle to the correct diagnosis of the phase of the system, but for this reason any figures presented displaying numerical calculations of the ground state wavefunction have been performed using $50\times50$ lattices.

\begin{figure}[htb]
\begin{center}
a)
\epsfig{file=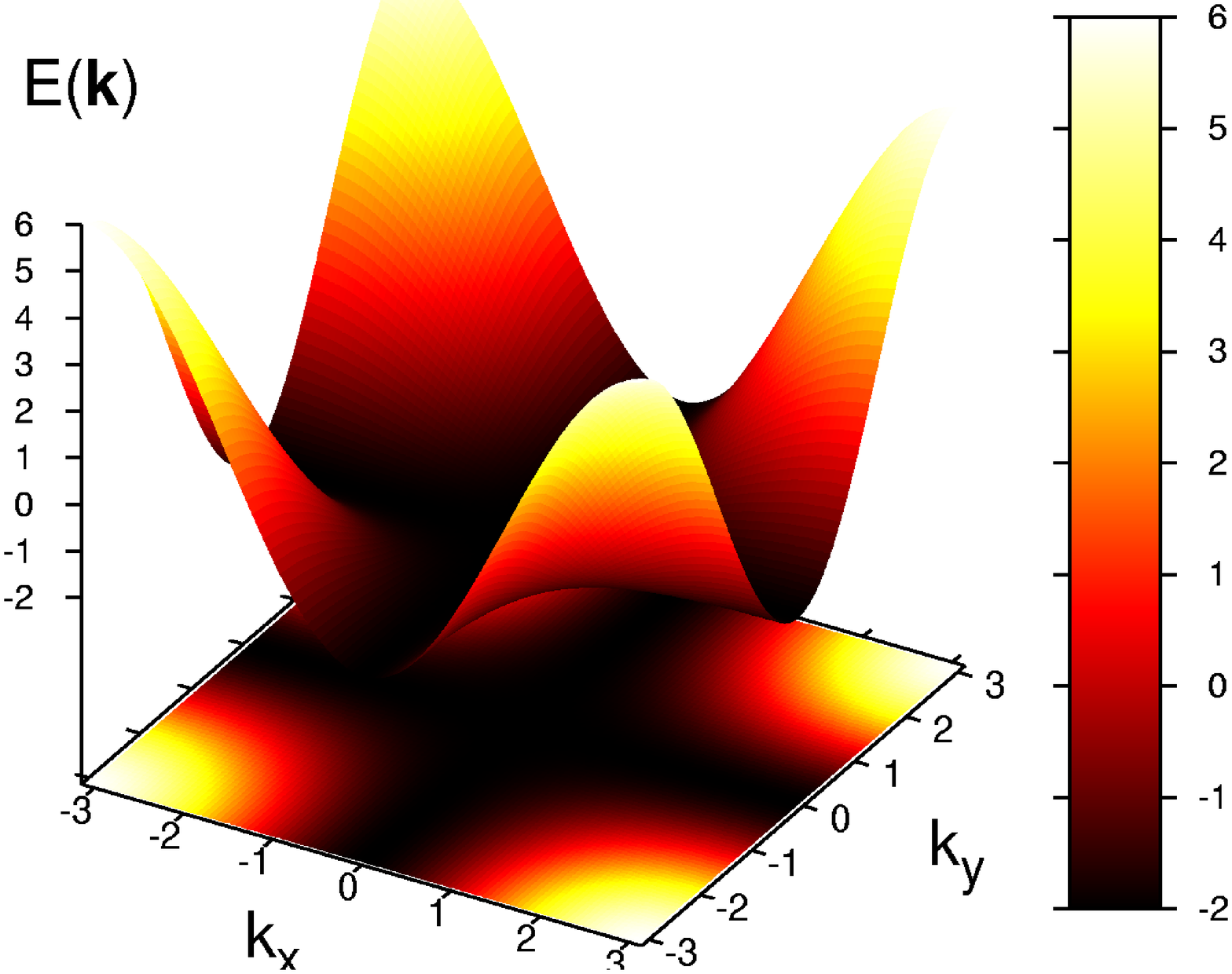,height=5.4cm}
\hspace{.5cm}
b)
\epsfig{file=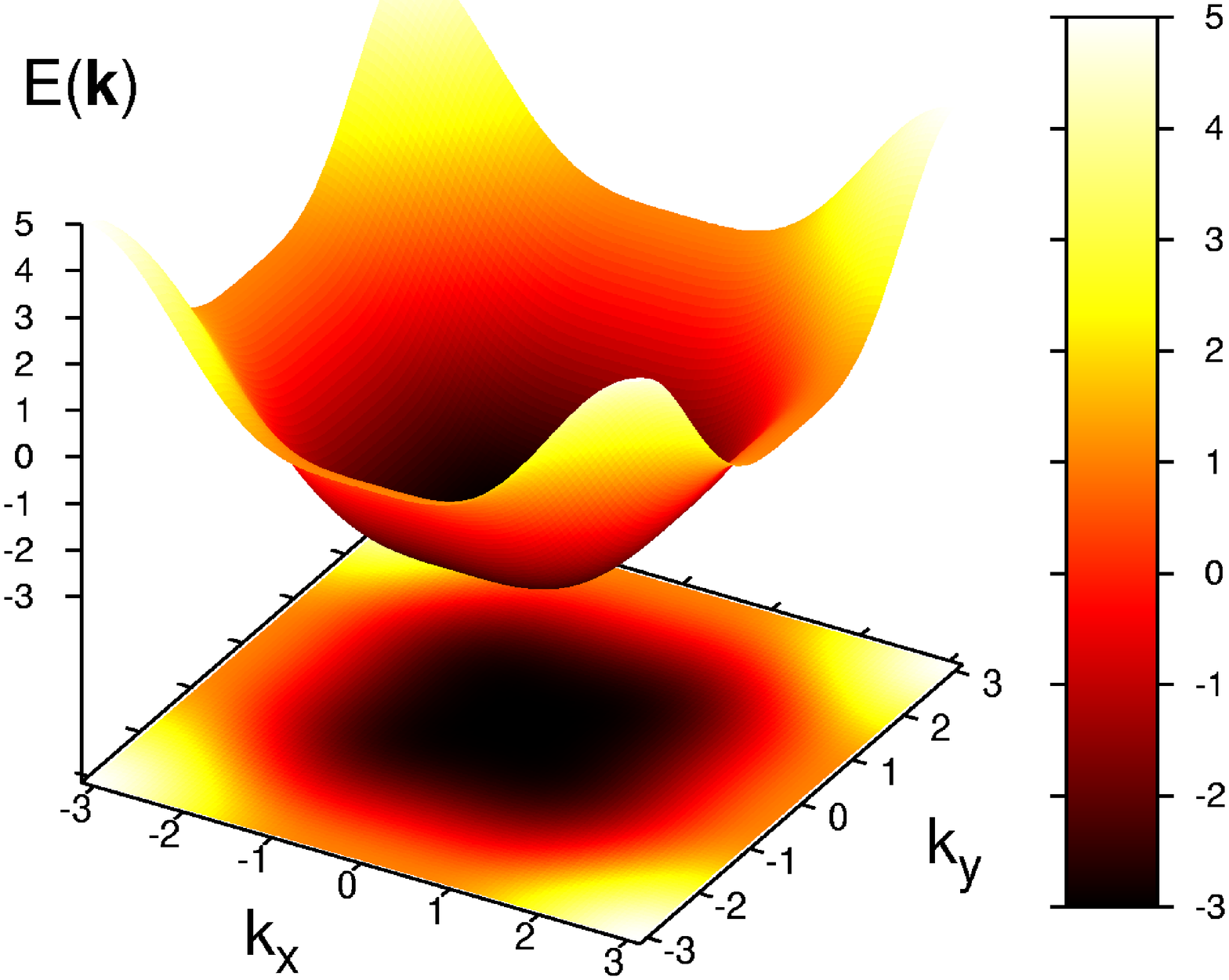,height=5.4cm}
\end{center}
\begin{center}
c)
\epsfig{file=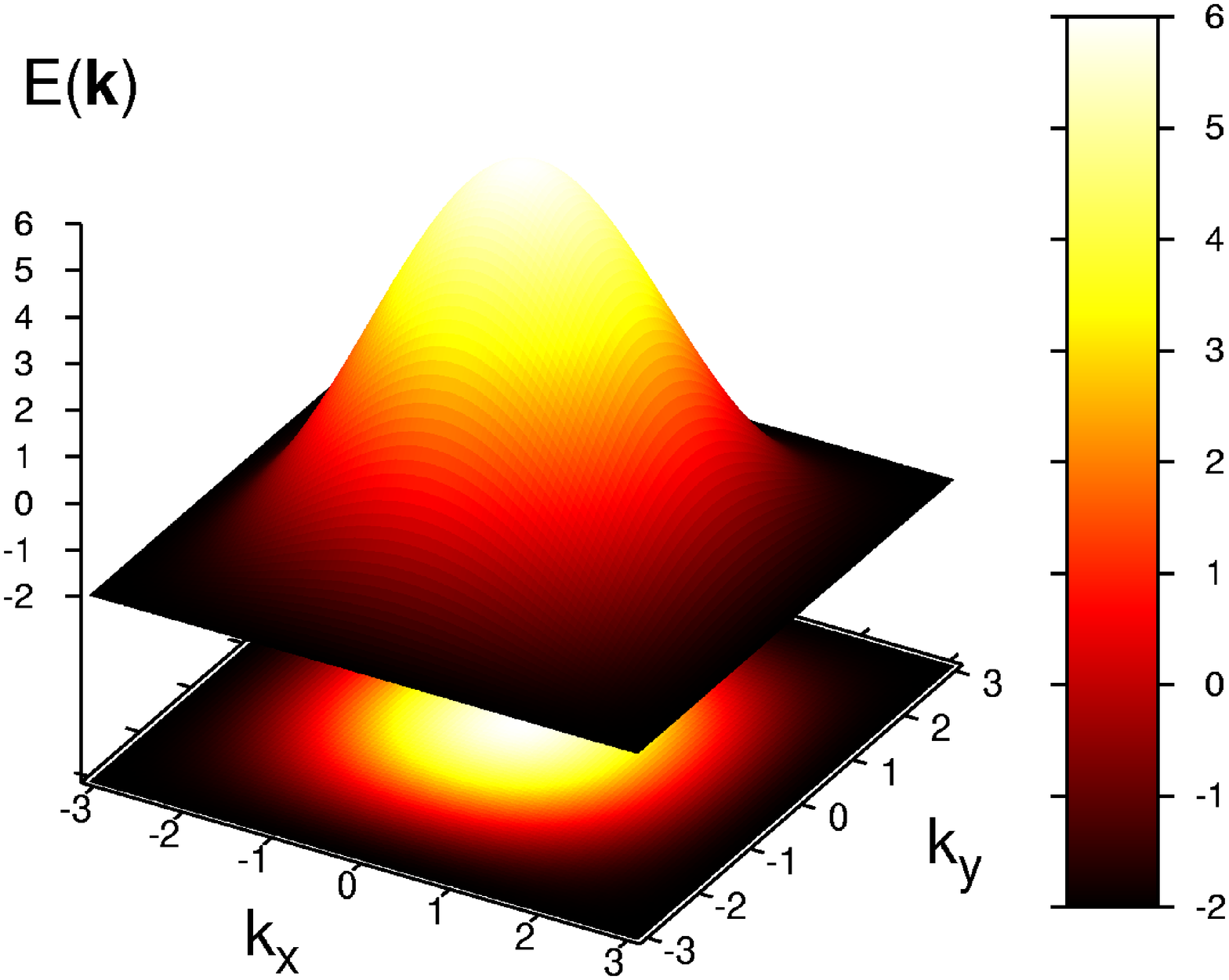,height=5.4cm}
\hspace{.5cm}
d)\epsfig{file=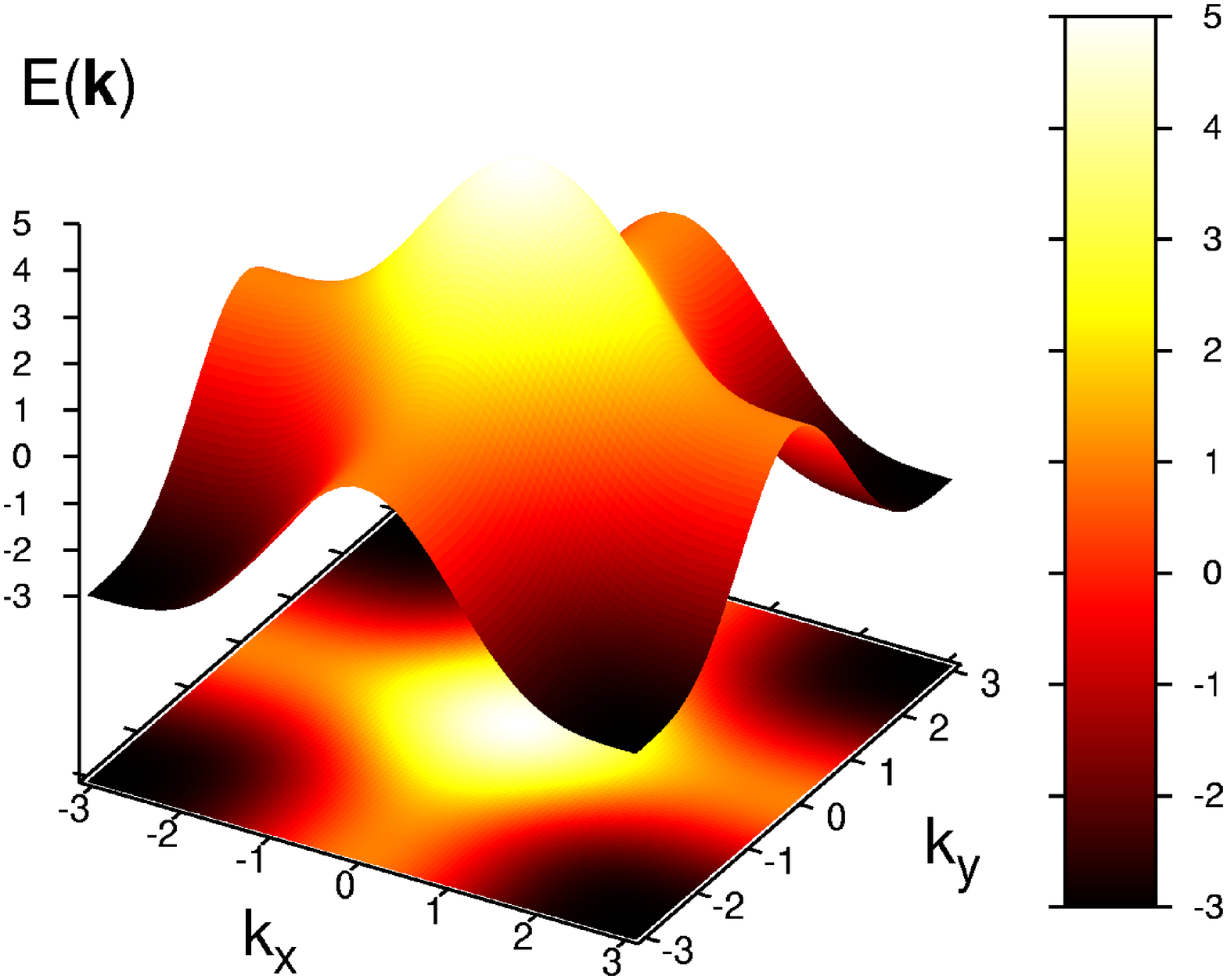,height=5.4cm}
\end{center}
\vspace{-.5cm}
\caption{\small Dispersion relations for a) $t_2=-0.5$ and b) $t_3=-0.25$ in the conventional case, and c) $t_2=-0.5$ and d) $t_3=-0.25$ in the unconventional case .  }
\label{fig:Denergy2}
\end{figure}


\section{Kinetic Energy Driven Phase Transitions}\label{sec:KPTout}

Provided that the repulsion is weak, the ground state of the system is determined by the form of the kinetic contribution to the Hamiltonian, which in the tight binding limit incorporates the effects of the periodic lattice into the values of the hopping parameters $t$:
\begin{equation}
E({\bf k})=4t_0-2t_1(\cos(k_x)+\cos(k_y)) - 4t_2\cos(k_x)\cos(k_y) 
-2t_3(\cos(2k_x)+\cos(2k_y)), \label{eqn:disp}
\end{equation}

\noindent where for the purposes of our present discussion we take $t_0=0$ (its precise value is of little relevance here).  The basic form of the band is given by the sign of $t_1$ (whose absolute value is taken to be unity here and thereafter): an conventional dispersion has $t_1$ positive and a single minimum  at the $\Gamma$ point and a unconventional dispersion has $t_1$ negative and four minima located in the corners of the Brillouin zone (Figure \ref{fig:Denergy1}).  At small values of the repulsion the condensate wavefunction in the former case is flat, in the latter case it has a staggered (alternating) sign with each lattice site.  In both cases the condensate density is homogeneous.

\begin{figure}[htb]
\begin{center}
a)
\epsfig{file=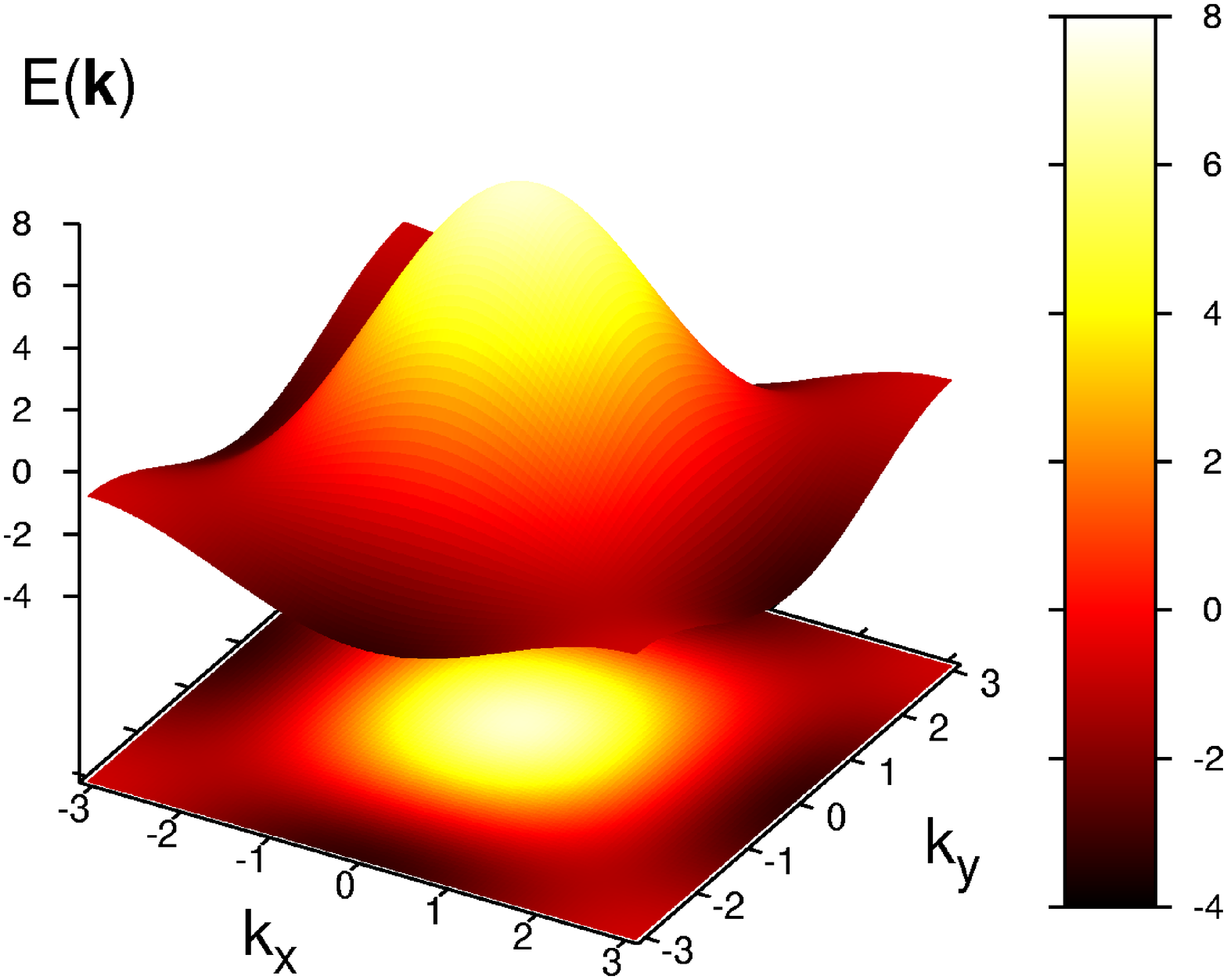,height=5.4cm}
\hspace{.3cm}
b)
\epsfig{file=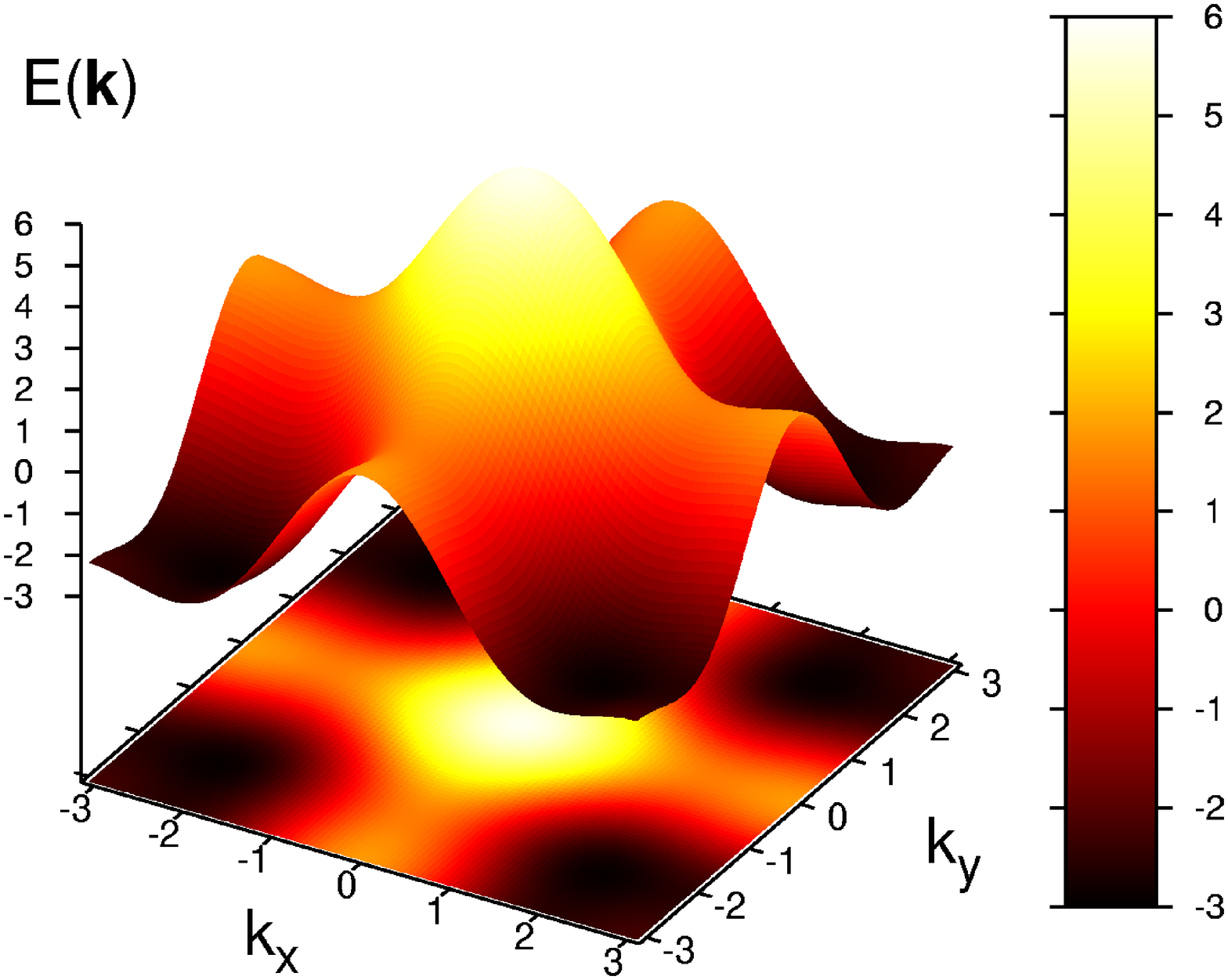,height=5.4cm}
\end{center}
\vspace{-.5cm}
\caption{\small Dispersion relations for the unconventional case with  a) $t_2=-0.8$ (C phase) and b) $t_3=-0.25$ (D phase).}
\label{fig:Denergy3}
\end{figure}

We restrict our discussion to cases where $t_2,t_3<0$, since this appears to be where many of the phenomena of interest reside.  As discussed in \cite{Alexandrov:2008}, we may modify the behaviour of the system by decreasing the values of the hopping parameters $t_2$ and $t_3$.  Varying $t_2$ only or $t_3$ only we discover that the value of the inverse effective mass is decreased until at certain critical values ($t^c_2\approx-0.5$ and $t^c_3\approx-0.25$) it becomes zero (this corresponds to a flattening of the dispersion in the vicinity of the minima -- see Figure \ref{fig:Denergy2}). As our discussion in \S\ref{sec:Intro} indicates, in the conventional case we should take these to be the lower limit of the allowed values of those hopping parameters, since beneath these values the ground state of the system develops nodes, and this is unphysical for the ground state of a bosonic system.  In the unconventional case, however, we take this to be an indication of a phase transition where the wave-vector/inverse period of the condensate modulation is the order parameter; here the nodes of the system change location.  When $t_2<t^c_2$, the minima are now located at at ${\bf k_1} = (\pi, 0)$, ${\bf k_2}= (-\pi, 0)$, ${\bf k_3}=(0,\pi)$ and ${\bf  k_4}=(0, -\pi)$; this is a first-order phase transition due to the discontinuous change in the location of the minima.  We shall call this the {\bf Commensurate phase} (C phase) due to the appearance of modulations in the ground state that are commensurate with the lattice period.  When $t_3<t_{3c}$, these are located at ${\bf k_1} = k(1, 1)$,${\bf k_2}= k(-1, 1)$, ${\bf k_3}=k(-1,-1)$ and ${\bf k_4}=k(1, -1)$, where $k$ is a constant factor determined by the value of $t_3$ that is equal to $\pi$ (unconventional case) or $0$ (conventional) at $t^c_3$ and tends towards $\pi/2$ as $t_3$ is decreased \cite{Alexandrov:2008}. This results in modulations in the ground state that are aligned diagonally but that are not commensurate with the lattice period: we therefore denote this as the {\bf Diagonal phase} (D phase).  Due to the continuous change in the order parameter, this is a second-order phase transition.  Examples of dispersion relations within these phases may be found in Figure \ref{fig:Denergy3}.  We call the phase transitions `kinetic energy driven' as it is the form of the tight-binding kinetic dispersion relation that drives them.

\begin{figure}
\begin{center}
a)
\epsfig{file=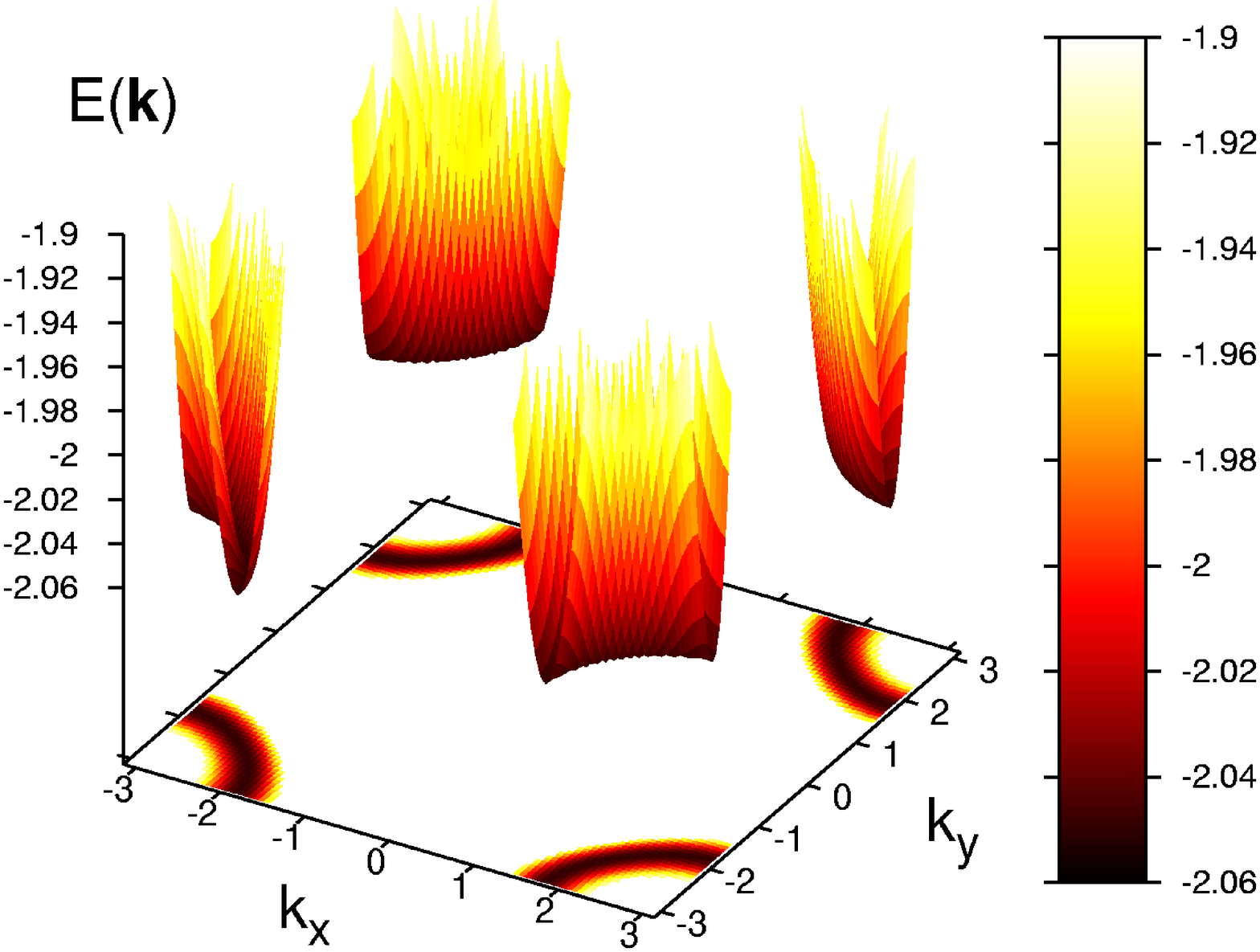,height=5cm}
\hspace{.3cm}
b)
\epsfig{file=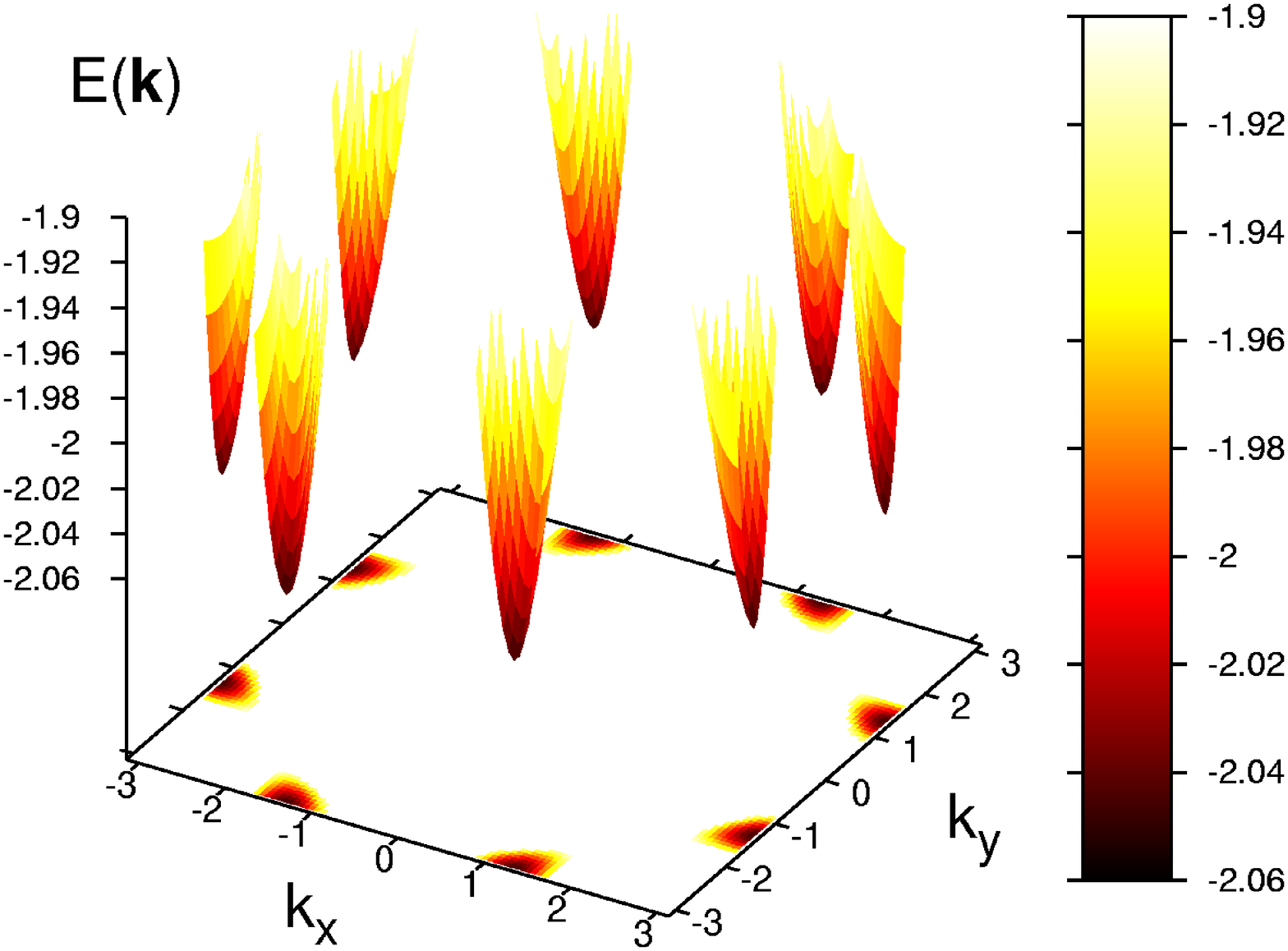,height=5cm}
\end{center}
\begin{center}
c)
\epsfig{file=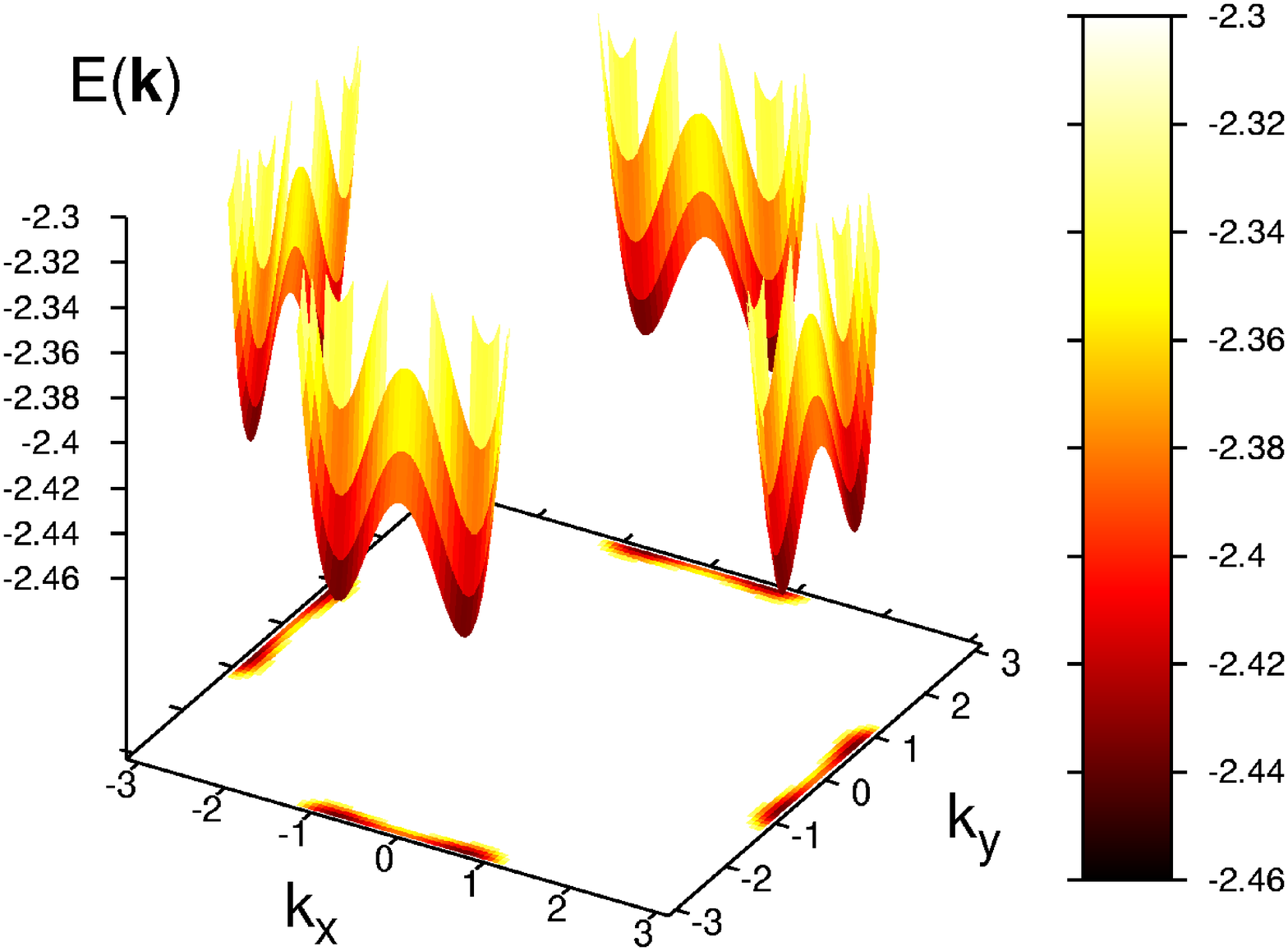,height=5cm}
\hspace{.3cm}
d)\epsfig{file=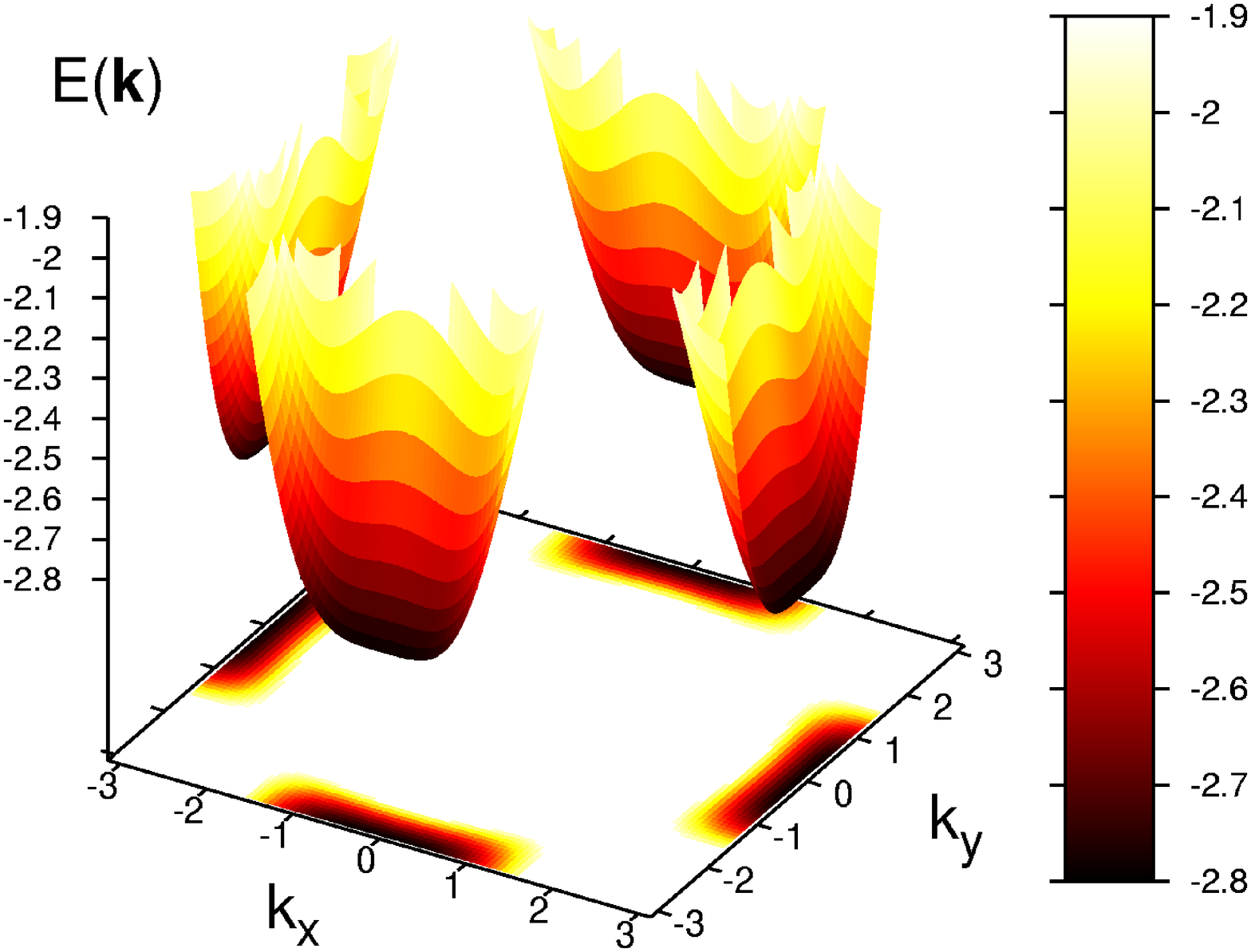,height=5cm}
\end{center}
\vspace{-.5cm}
\caption{\small Low-energy portions of dispersions in the unconventional case with $t_3=-0.2$ and a) $t_2=-0.4$, b) $t_3=-0.5$, c) $t_2=-0.8$ and d) $t_2=-0.9$, showing the rotation of the minima.}
\label{fig:Denergy4}
\end{figure}

The situation when both $t_2$ and $t_3$ are varied together is somewhat more complex, as varying both has an effect on the location of the transitions to the various phases.  In addition, we may see that there exists an additional phase.  In the conventional case, we have four minima located at ${\bf k_1} = (l,0)$, ${\bf k_2}= -(l,0)$, ${\bf k_3}=(0,l)$ and ${\bf  k_4}=\pi(0,l)$ where $l$ is dependent on the values of the hopping parameters and tends towards $\pi$ if $t_2$ alone is increased.  The unconventional case is equivalent to a displacement of the centre of the conventional Brillouin zone by $(\pi/2,\pi/2)$ and so there are {\em eight} minima instead of four, located at ${\bf k^\pm_1} = \pi(1,\pm l)$, ${\bf k^\pm_2}= -\pi(1,\pm l)$, ${\bf k^\pm_3}=\pi(\pm l,1)$ and ${\bf  k^\pm_4}=-\pi(\pm l,1)$, where $l$ is again dependent on the values of the hopping parameters and tends towards $0$ if $t_2$ alone is increased.   Figure \ref{fig:Denergy4} shows the evolution of the wave vectors for $t_3=-0.2$ as $t_2$ is varied -- in this case, the wave-vectors are rotated away from the diagonal axis of the D phase, until they finally reach the vertical and horizontal axes of the C phase.  We therefore designate this to be the {\bf Rotated phase} (R phase).

\begin{figure}[htb]
\begin{center}
\epsfig{file=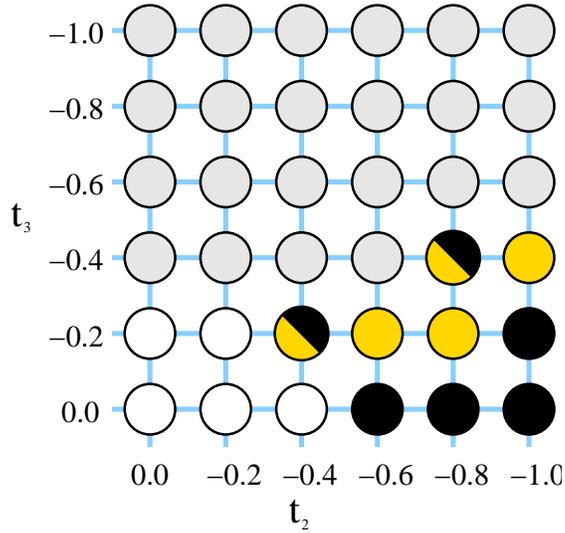,height=7cm}
\end{center}
\vspace{-.5cm}
\caption{\small The form of the dispersion on the $t_2-t_3$ plane.  White signifies the ordinary homogeneous phase, gray the D phase, black the C-phase and gold the R-phase.  A divided circle indicates a dispersion intermediate between two phases, as in Figure \ref{fig:Denergy4} a).}
\label{fig:Phase1}
\end{figure}

Examination of the form of the dispersion relation for different values of $t_2$ and $t_3$ allows us to classify various regions of the parameter space according to these phases.  This is done in Figure \ref{fig:Phase1}.

One method of realising this behaviour in optical lattices is through the use of shallow lattice potentials, such that the  localisation of the Wannier functions describing the lattice is poor. Stojanovi\'c {\em et al.} \cite{Stojanovic:2008} predict incommensurate condensate wave vectors for atoms in the meta-stable p-band of a double-well lattice  for sufficiently small values of the lattice potential; the example given of a corresponding dispersion indicates that this is within the region of our D-phase. Note that in most cases involving optical lattices, conventional type dispersions with more than one zero are likely unphysical if we are in the s-wave ground state; in this case, Figure \ref{fig:Phase1} displays possible bounds on the allowed values of $t_2$ and $t_3$.

In the next sections, we shall discuss the effects of the presence of different forms of potential on the condensate wave function for different values of the hopping parameters.

\begin{figure}[htb]
\begin{center}
\epsfig{file=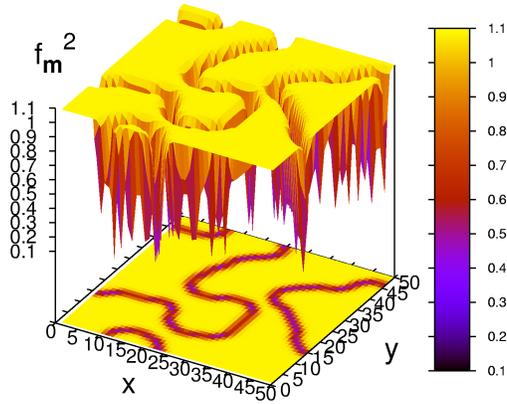,height=5.4cm}
\end{center}
\caption{\small Inhomogeneity in the density for the unconventional case when {$u_{00}=4.0$} and additional hopping terms are zero.}
\label{fig:inhom1}
\end{figure}

\section{Intermediate values of $u_{00}$ and additional hopping parameters}\label{sec:swallowtail}

What can be considered to be intermediate values of $u_{00}$?  In  Table \ref{table:SF-Mott}, we show various critical values of $u^c_{00}$ resulting from numerical simulations of the Bose-Hubbard model on a 2D square lattice with commensurate density.  In what follows, we take `intermediate' to mean any reasonably large ($u_{00}>1)$ value of the repulsion which is also below $u^c_{00}$, and is therefore within the range of validity of the GPE.

\begin{table}
\begin{tabular}{|c|c|}
\hline
Simulation Type & Approximate $u^c_{00}$\\
\hline
Gutzwiller-type Monte Carlo\cite{Krauth:1992} & 23.32\\
Extended Gutzwiller-type Monte Carlo\cite{Cappello:2007,Cappello:2008} & 20.6\\
Green's Function Monte Carlo\cite{Krauth:1991,Capogrosso-Sansone:2008} & 16.7-17.0\\
\hline
\end{tabular}
\caption{Numerical results showing the location of the superfluid-Mott insulator transition for the Bose-Hubbard model at an average atomic density per site of {$n=1$} on a square, two-dimensional lattice.}
\label{table:SF-Mott}
\end{table}

For sufficiently large (but still intermediate in the above sense) values of the onsite repulsion $u_{00}$, we observe that the homogeneous ground-state becomes inhomogeneous (Figure \ref{fig:inhom1}), with long, meandering domain walls.  This is likely due to the repulsion becoming dominant over the periodic potential of the lattice, which in the tight-binding limit is incorporated into the kinetic portion of our Hamiltonian.  This effect in a bosonic system is reminiscient of a similar effect observed in fermionic systems \cite{Pang:1993}.

\begin{figure}[htb]
\begin{center}
a)
\epsfig{file=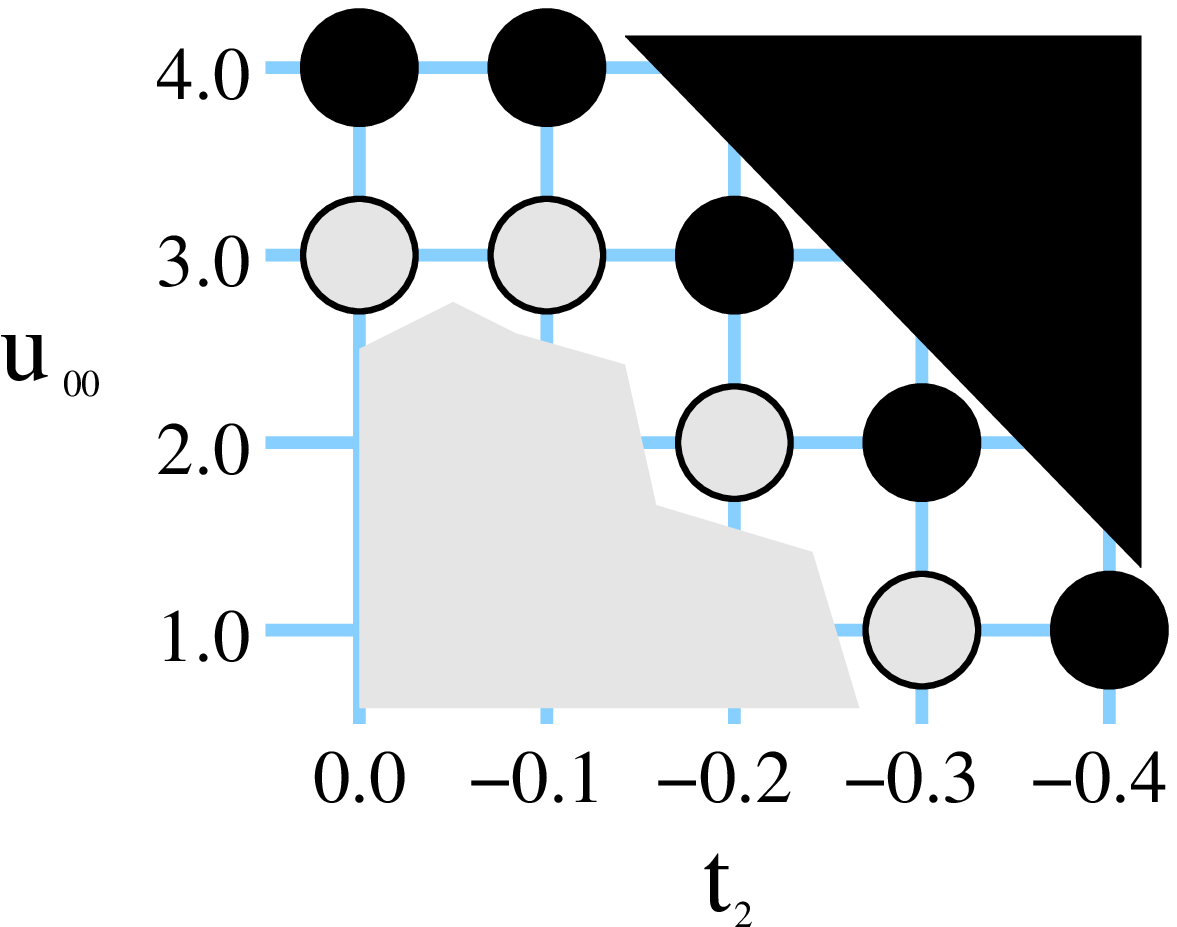,height=5.4cm}
\end{center}
\begin{center}
b)
\epsfig{file=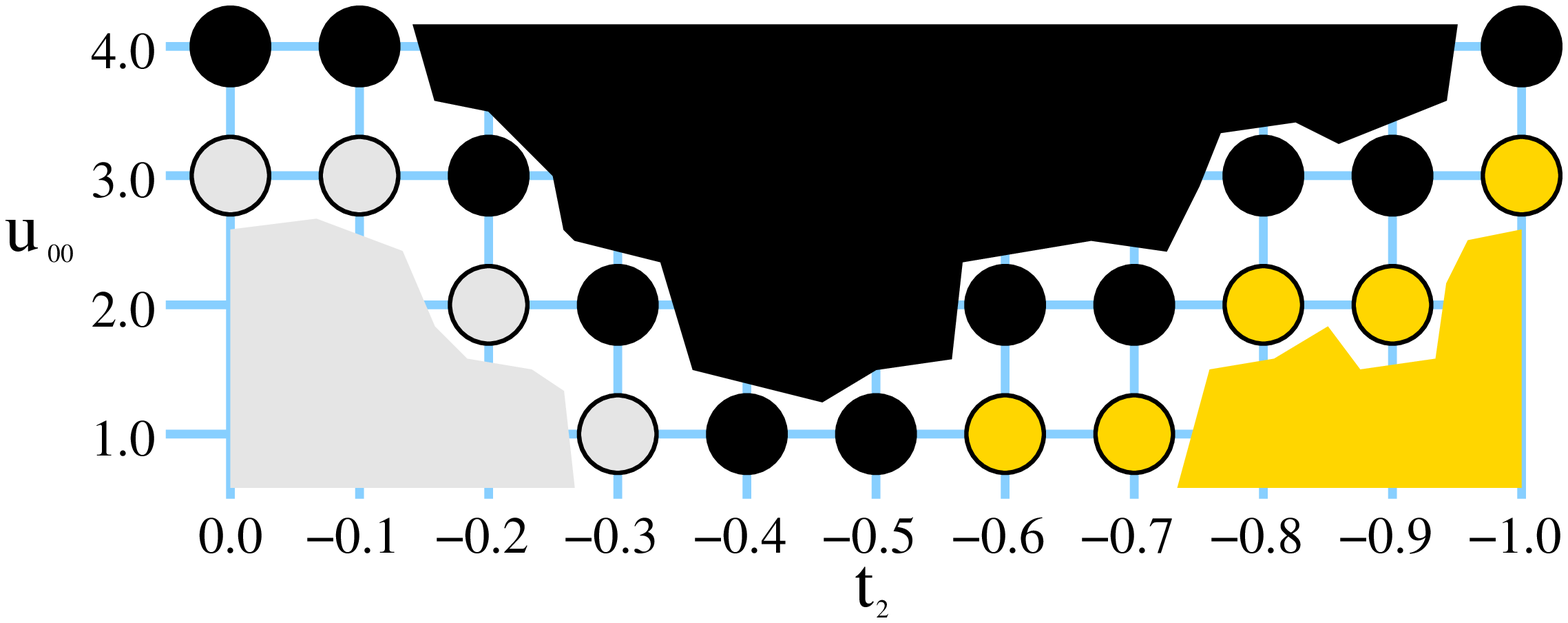,height=5.4cm}
\end{center}
\vspace{-.5cm}
\caption{\small Onset of inhomogeneity as $u_{00}$ and $t_2$ are varied.  The gray regions are homogeneous, the black inhomogeneous, and gold indicates the C phase, which has homogeneous density.  Circles correspond to simulations whose resulting ground state is given by their color. a) displays the conventional case when restricted to hopping values giving a single zero (taken from \cite{Alexandrov:2008}), and b) displays the unconventional case.}
\label{fig:Swallow1}
\end{figure}

An interesting feature of our model is the manner in which the critical value of the on-site repulsion at which this instability appears, $u^{in}_{00}$, is altered by the value of $t_2$. We have mapped out the approximate location of this transition through numerical simulation on $50\times50$ lattices for various values of $u_{00}$ and $t_2$, with $t_3$ and the other repulsions set to zero; our results are displayed in Figure \ref{fig:Swallow1}. For the conventional (first discussed in \cite{Alexandrov:2008})  and unconventional cases, we see that $u^{in}_{00}$ decreases as $t_2$ decreases from $0$ towards $-0.5$, which correlates with the decrease in depth of the minimum of the dispersion relation.  At $t_2=-0.5$, it is likely that $u^{in}_{00}=0$.  For the conventional case, decreasing $t_2$ further is normally unphysical, but for unconventional bosons this is not an issue; and for $t_2<-0.5$ we see that $u^{in}_{00}$ begins to increase once more until it returns to its original value at $t_2=-1$.

\begin{figure}
\begin{center}
a)
\epsfig{file=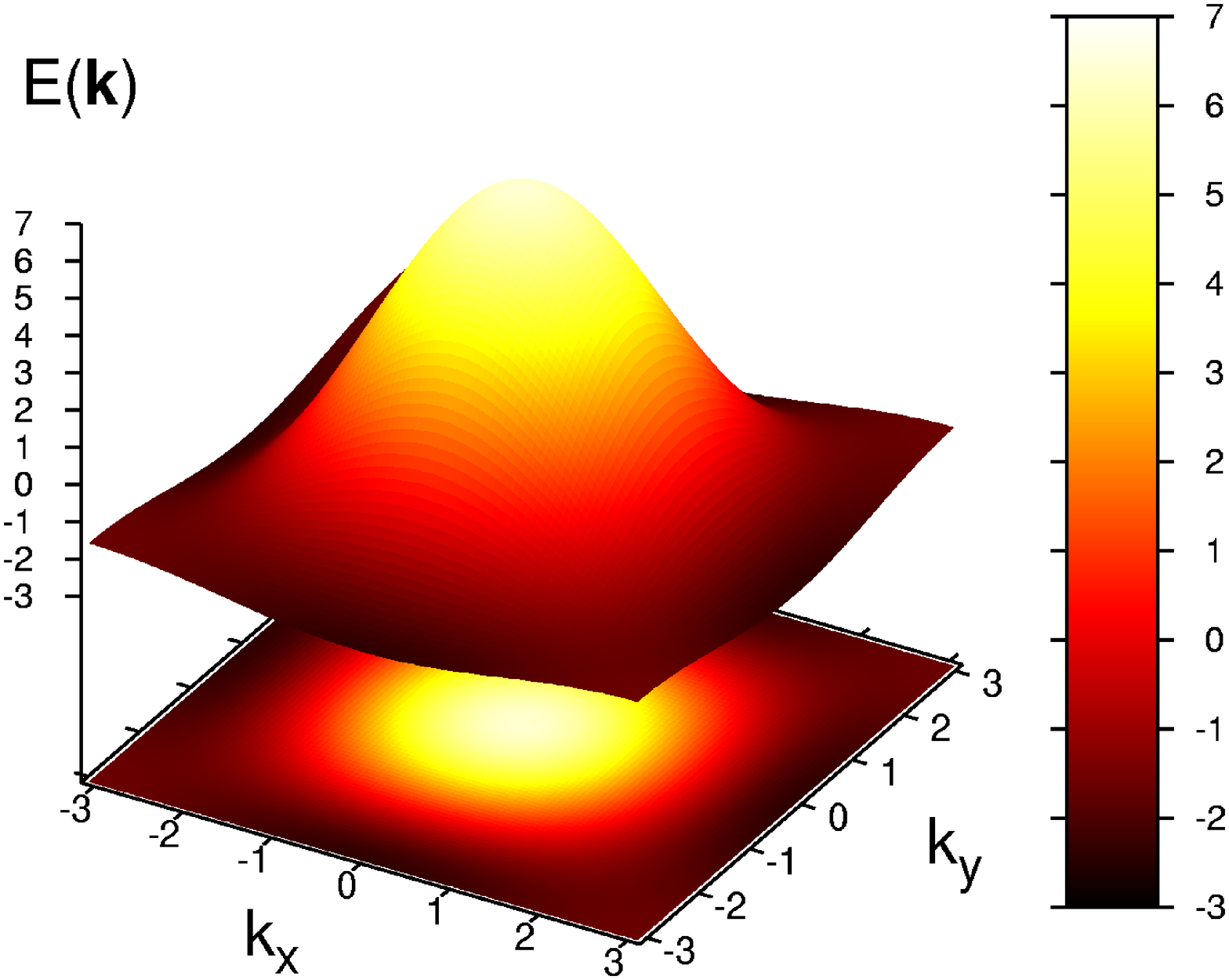,height=4.7cm}
\end{center}
\begin{center}
b)
\epsfig{file=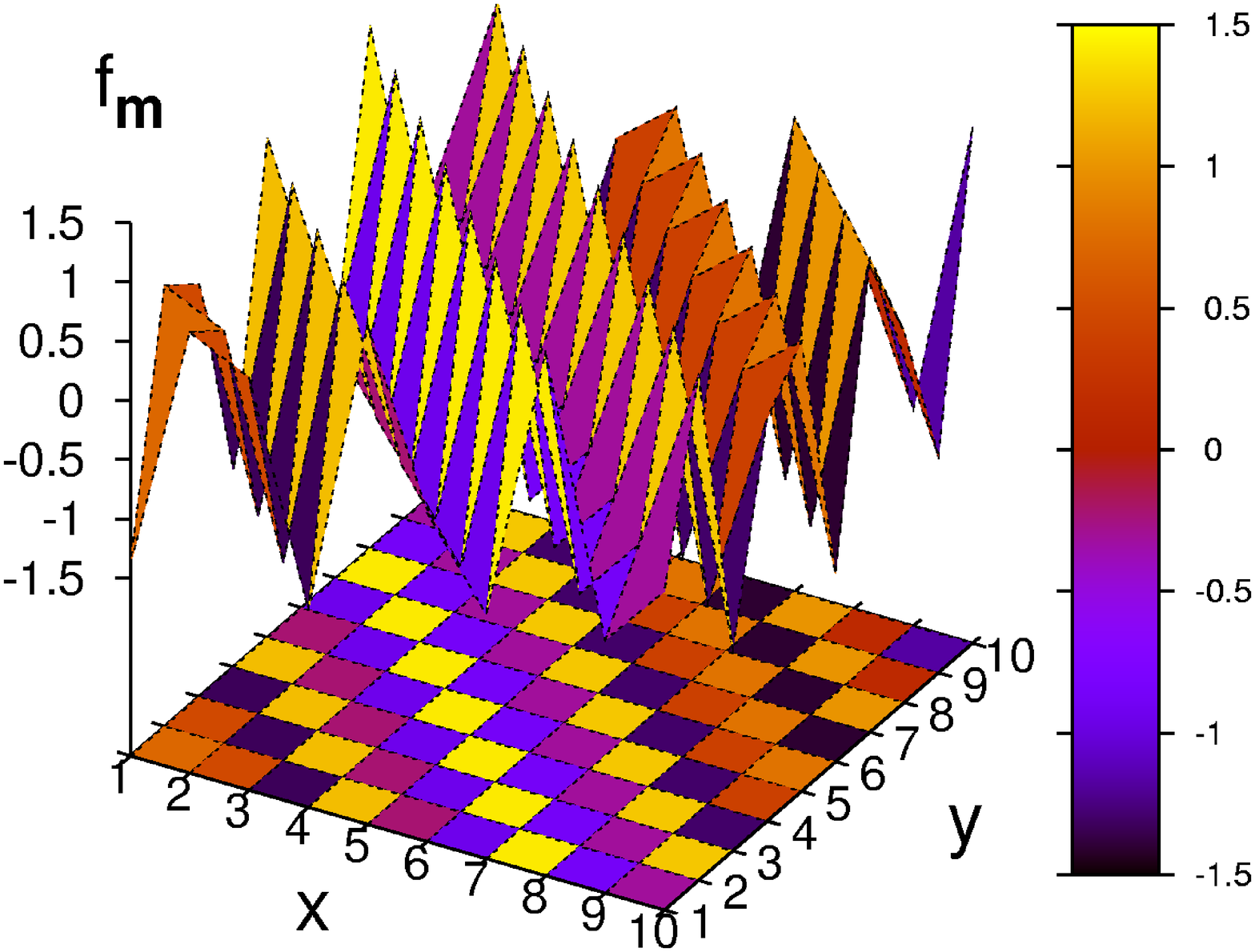,height=4.7cm}
\hspace{.3cm}
c)\epsfig{file=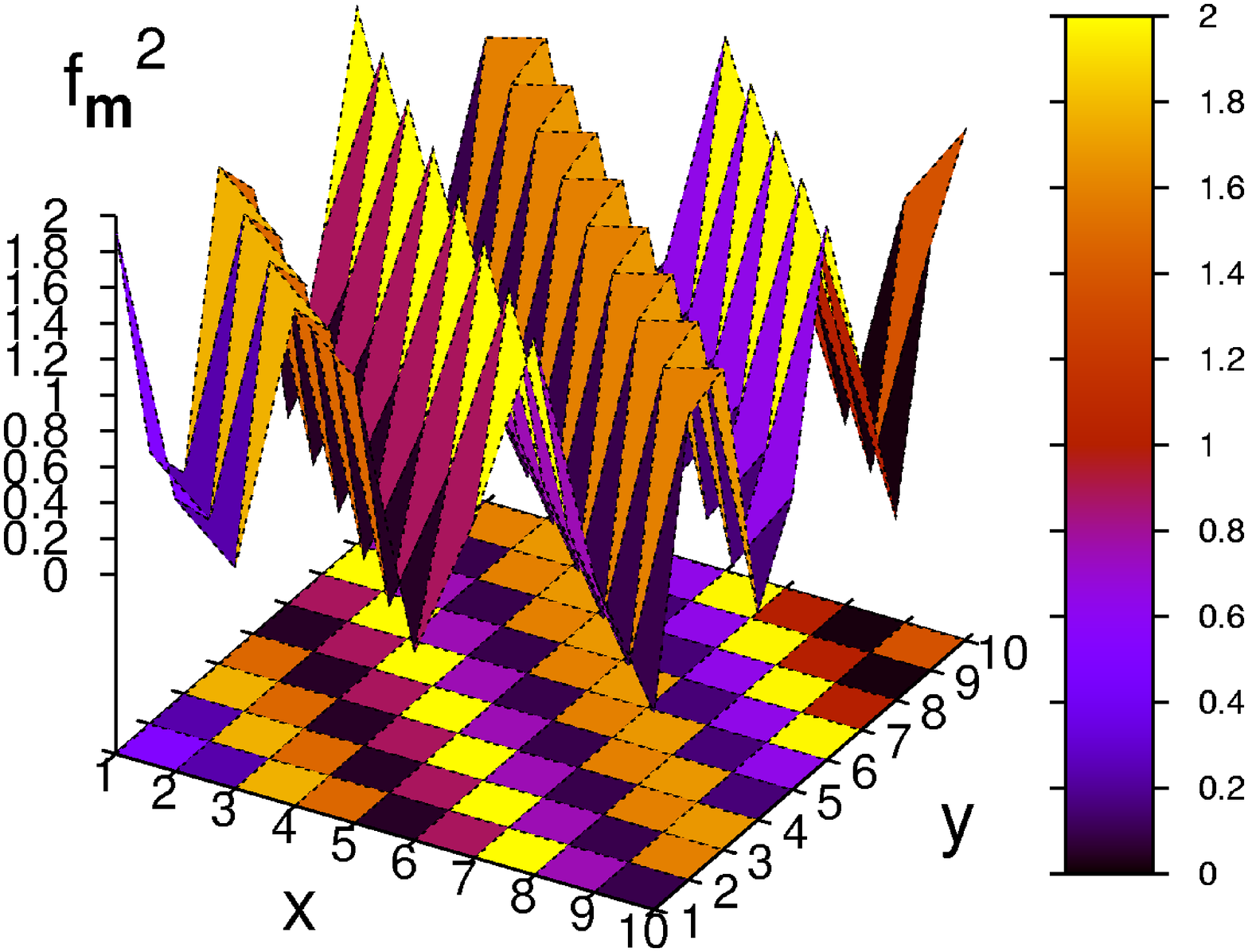,height=4.7cm}
\end{center}
\begin{center}
d)
\epsfig{file=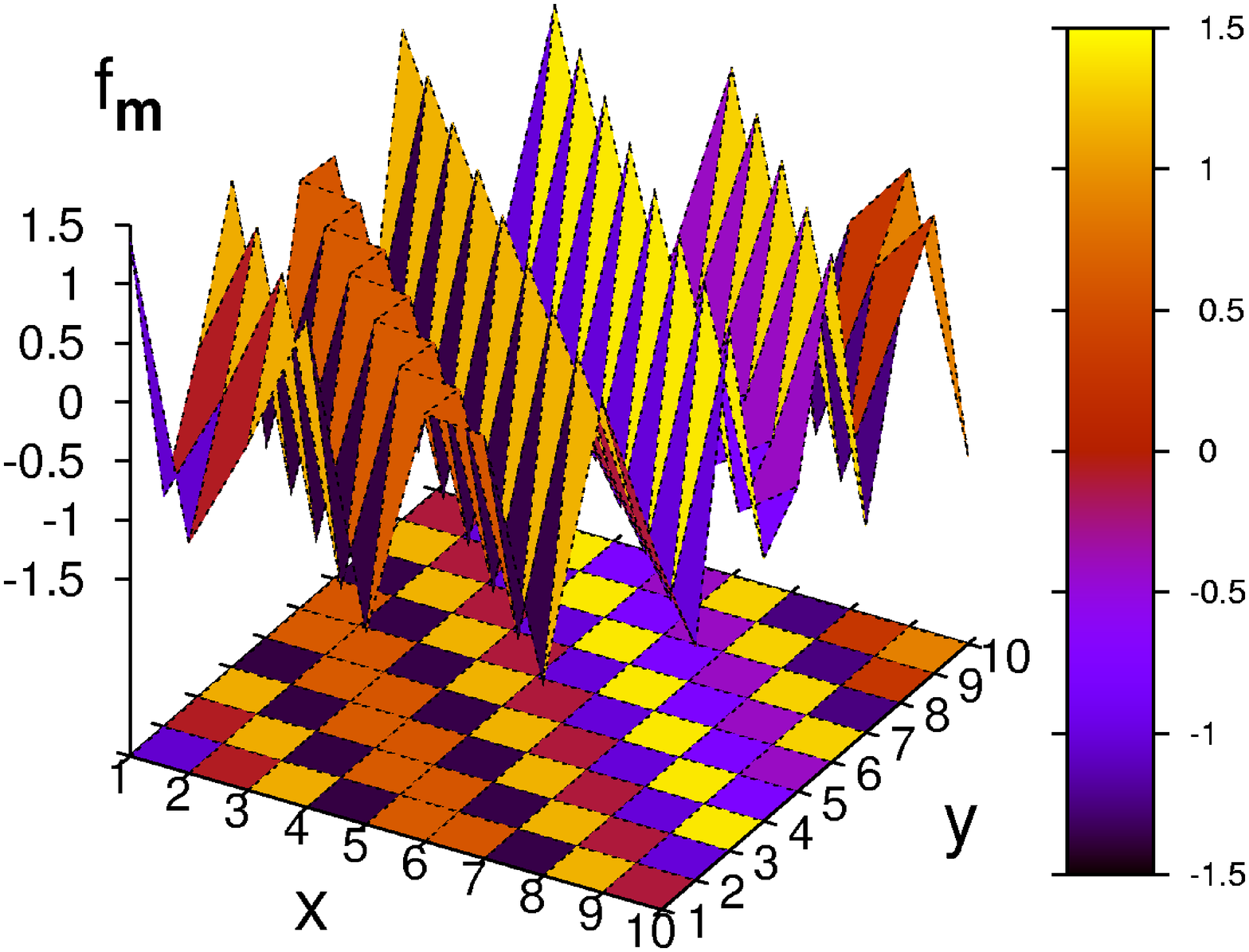,height=4.7cm}
\hspace{.3cm}
e)\epsfig{file=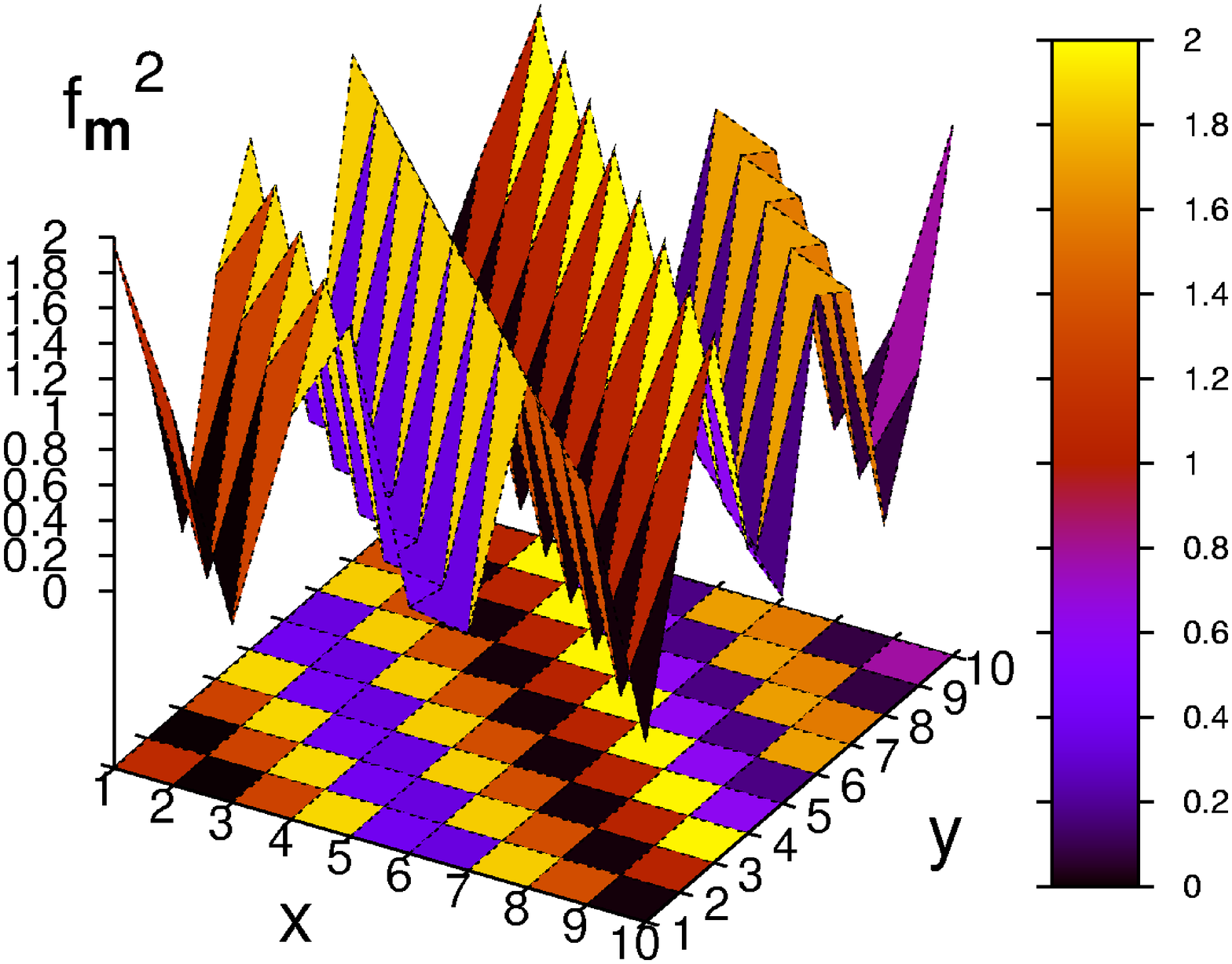,height=4.7cm}
\end{center}
\begin{center}
f)
\epsfig{file=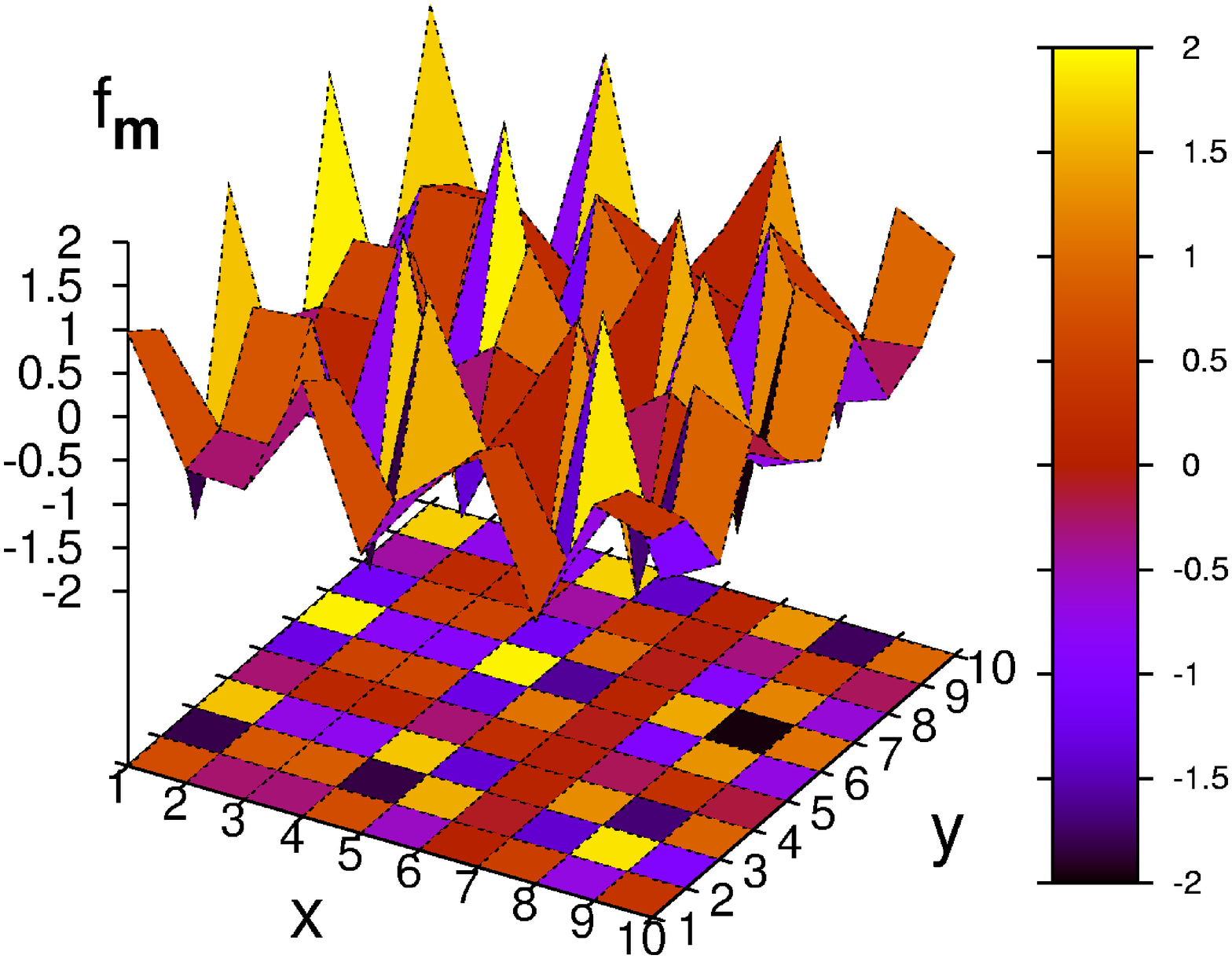,height=4.7cm}
\hspace{.3cm}
g)\epsfig{file=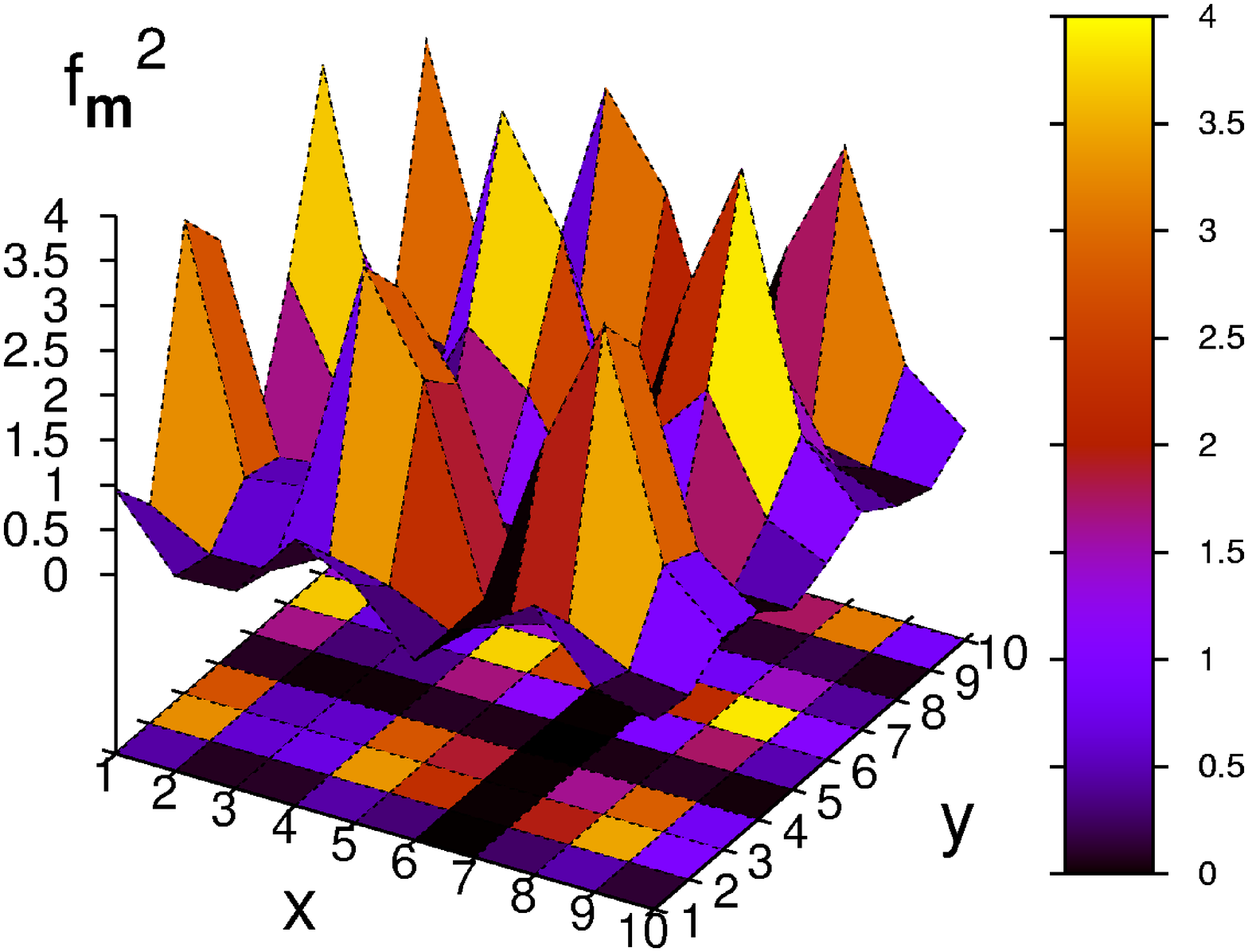,height=4.7cm}
\end{center}
\vspace{-.5cm}
\caption{\small Plots of a) the dispersion and the condensate wavefunctions and densities respectively for b) and c) $u_{00}=0.02$, d) and e) $u_{00}=0.02$ and $u_{01}=0.01$, f) and g) $u_{00}=0.02$ and $u_{02}=0.02$  in the unconventional case where $t_2=0$ and $t_3=-0.4$.  Colours are assigned to contour plots b)-g) according to value of the far corner of a square.}
\label{fig:Dcond1}
\end{figure}

The behaviour of $u^{in}_{00}$ as $t_2$ is varied corresponds closely with the height of the potential wall that separates one minimum from another (in the conventional case, that would be the height of the potential that separates the minimum at the $\Gamma$ point of the first Brillouin zone from the minima at the $\Gamma$ points of the neighbouring Brillouin zones).  As $t_2$ is decreased, the height of this potential wall is also decreased, until it becomes flat at $t_2=-0.5$, as in  Figures \ref{fig:Denergy2} a) and b).  As the height of the wall is decreased, the easier it is for the non-linear repulsion to overcome the kinetic energy.

\begin{figure}
\begin{center}
a)
\epsfig{file=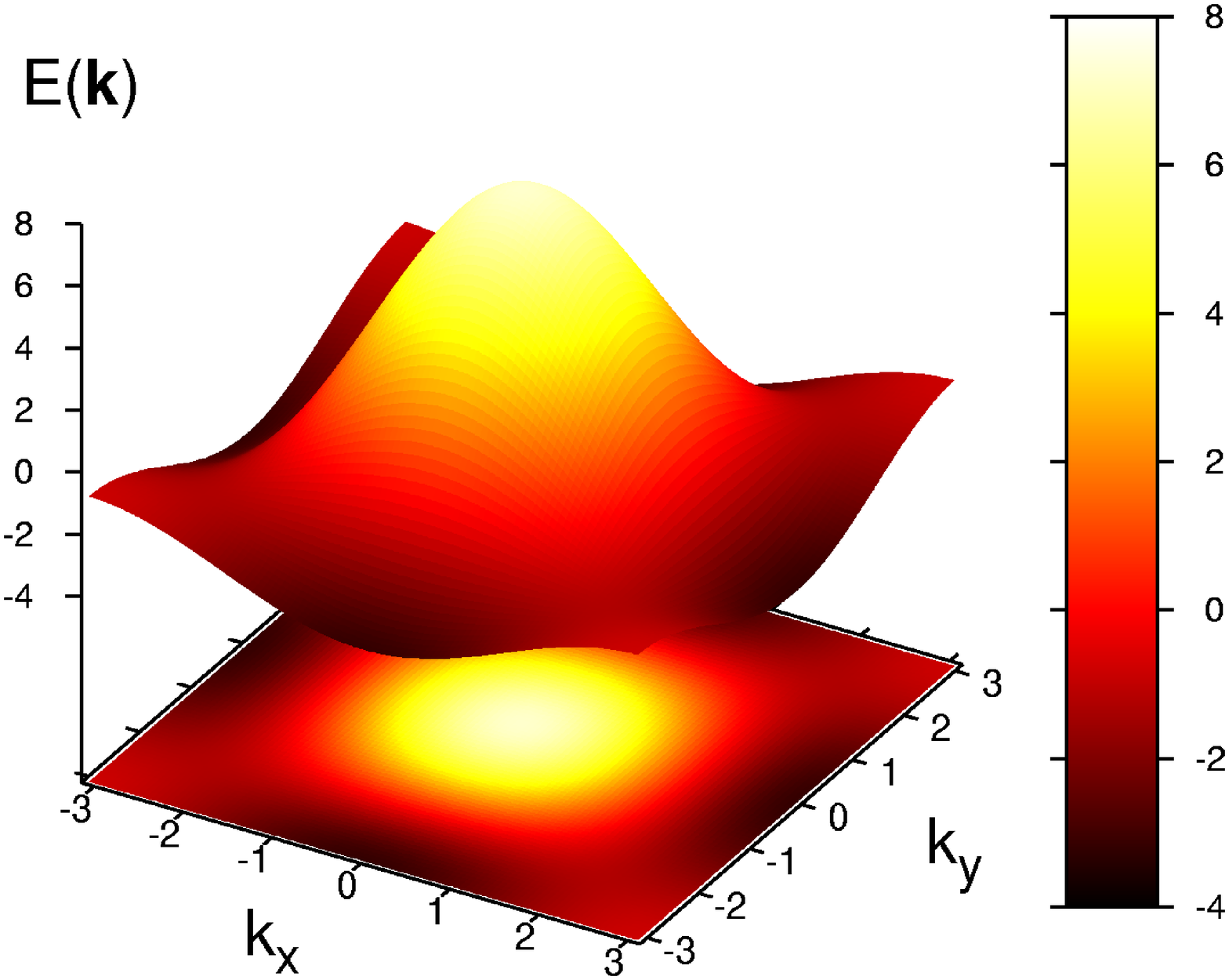,height=4.7cm}
\end{center}
\begin{center}
b)
\epsfig{file=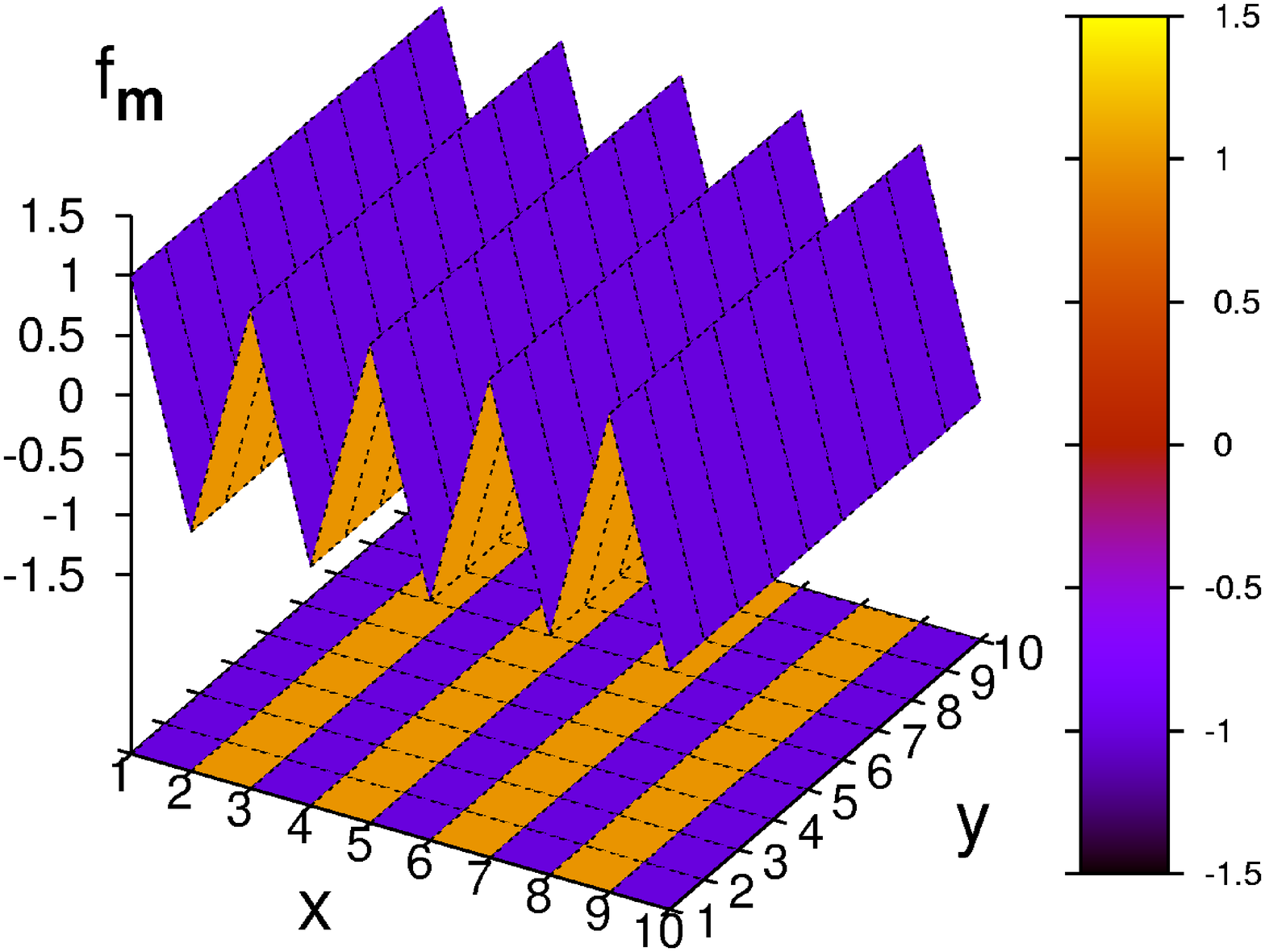,height=4.7cm}
\hspace{.3cm}
c)\epsfig{file=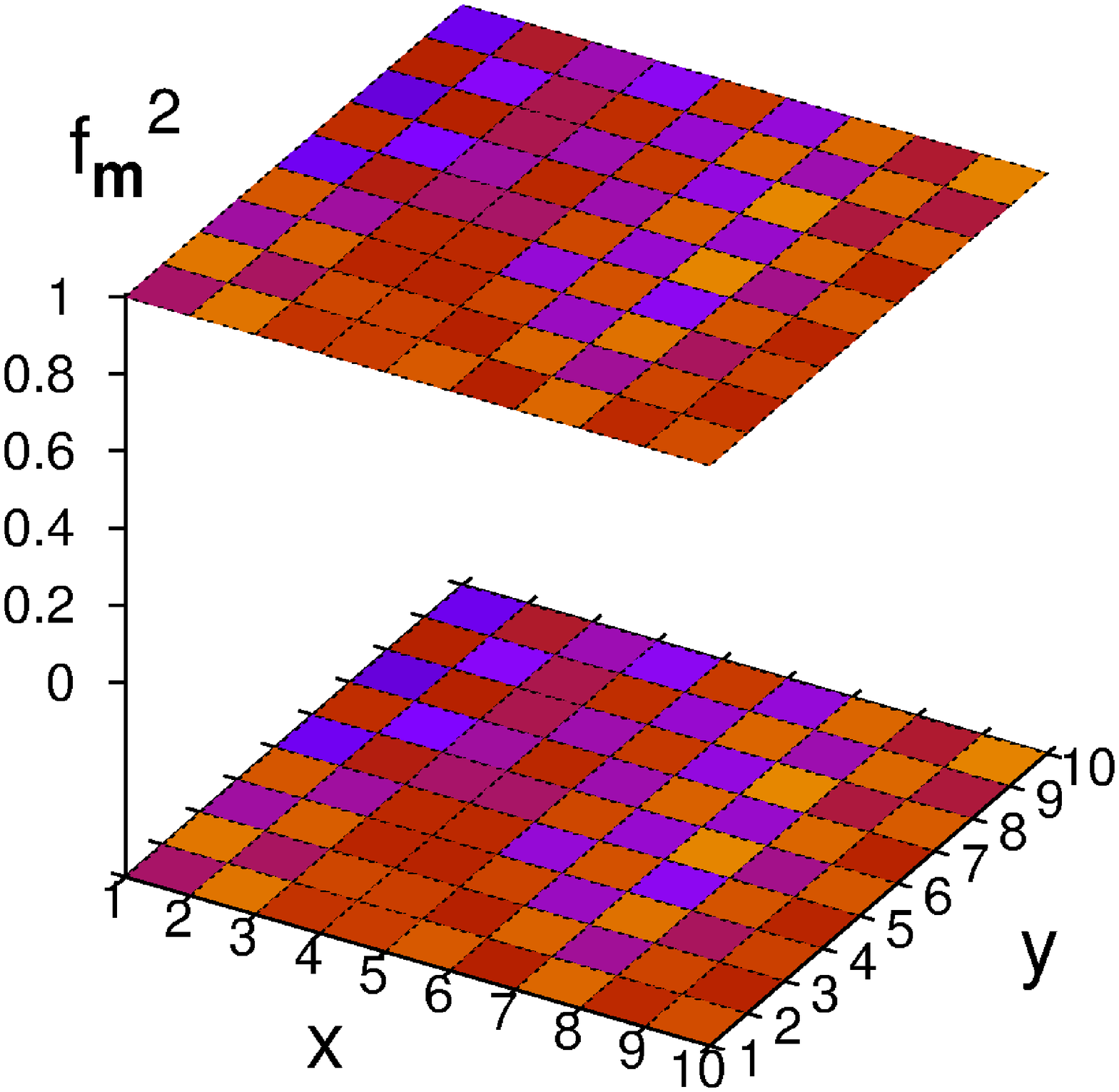,height=4.7cm}
\end{center}
\begin{center}
d)
\epsfig{file=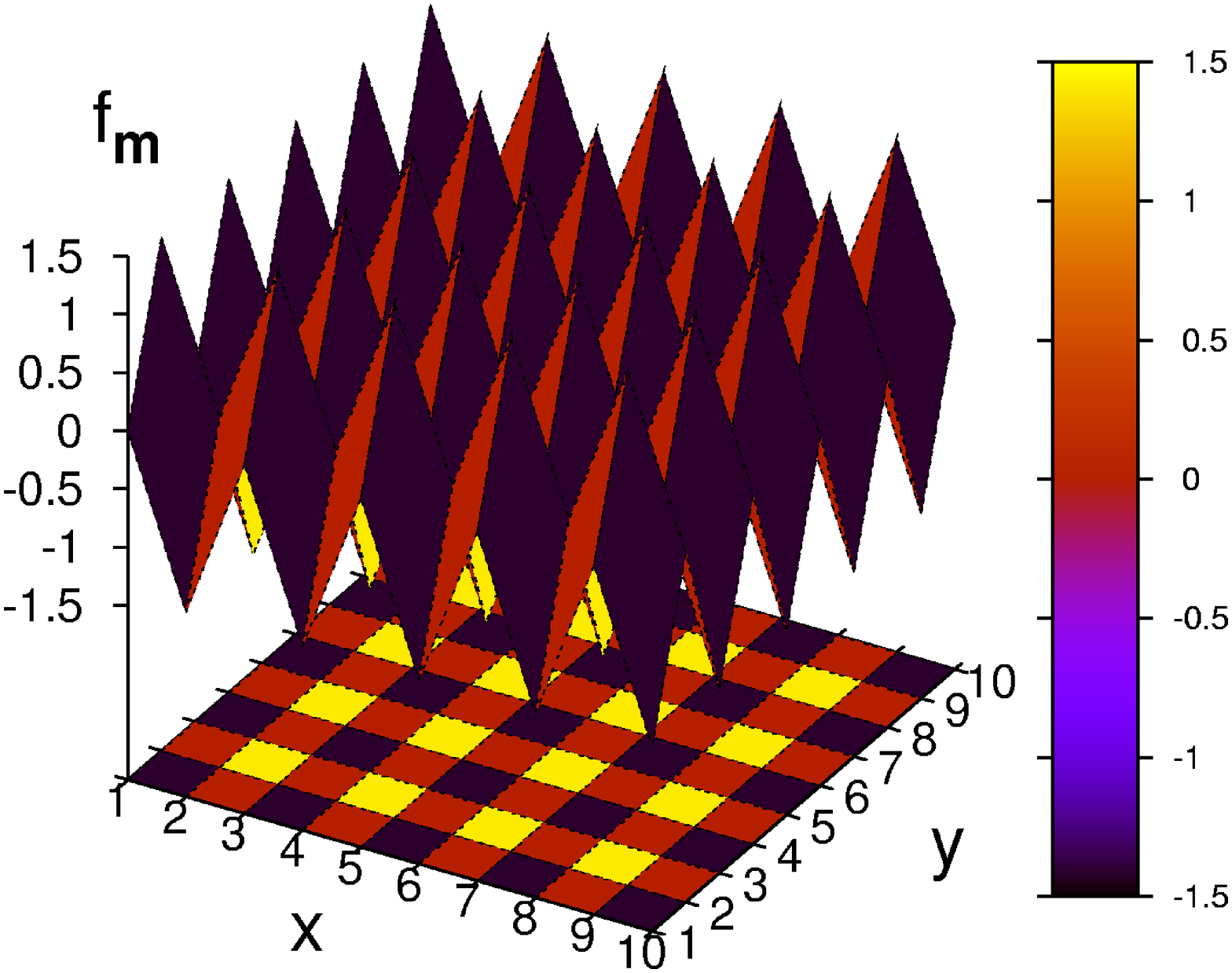,height=4.7cm}
\hspace{.3cm}
e)\epsfig{file=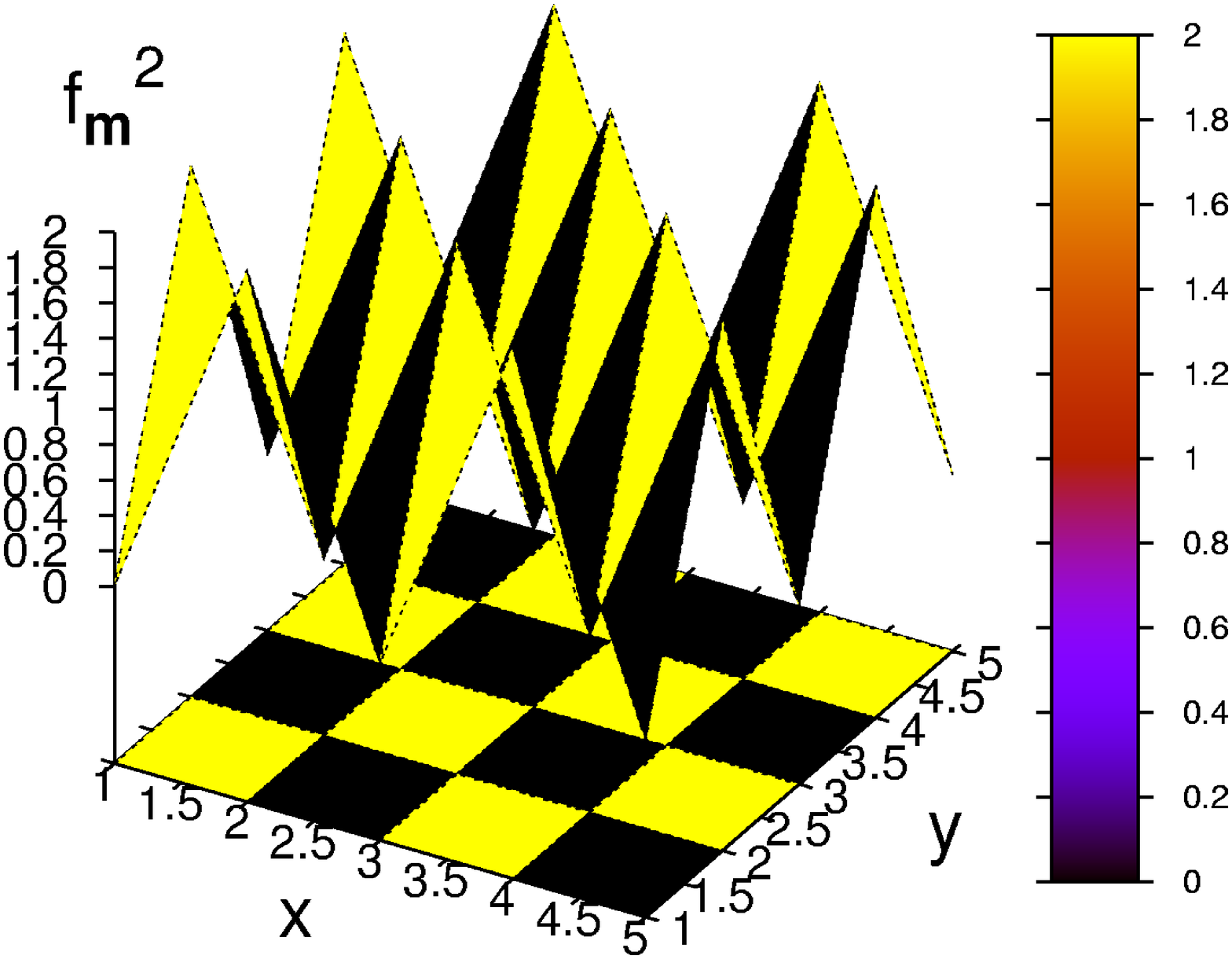,height=4.7cm}
\end{center}
\begin{center}
f)
\epsfig{file=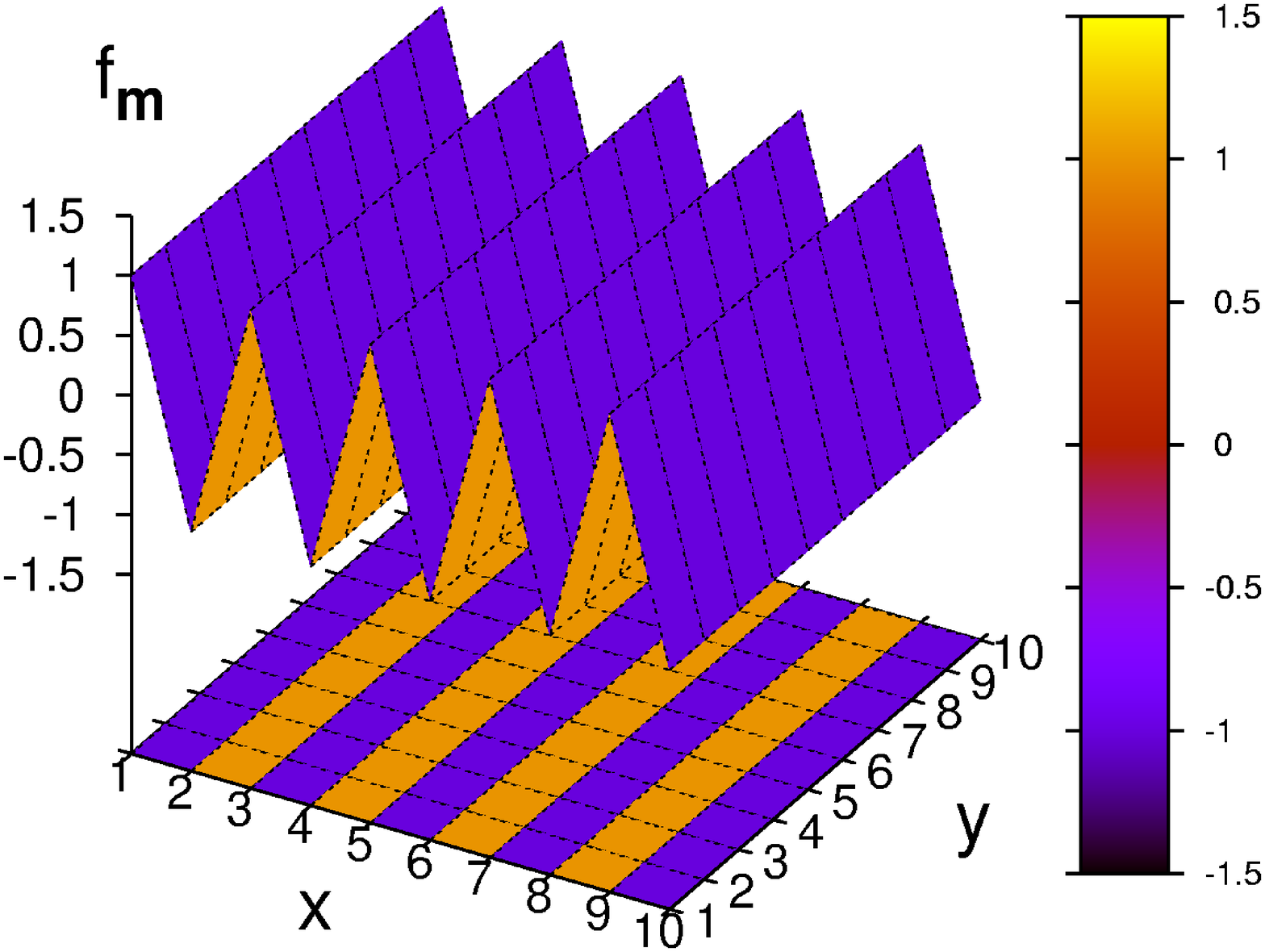,height=4.7cm}
\hspace{.3cm}
g)\epsfig{file=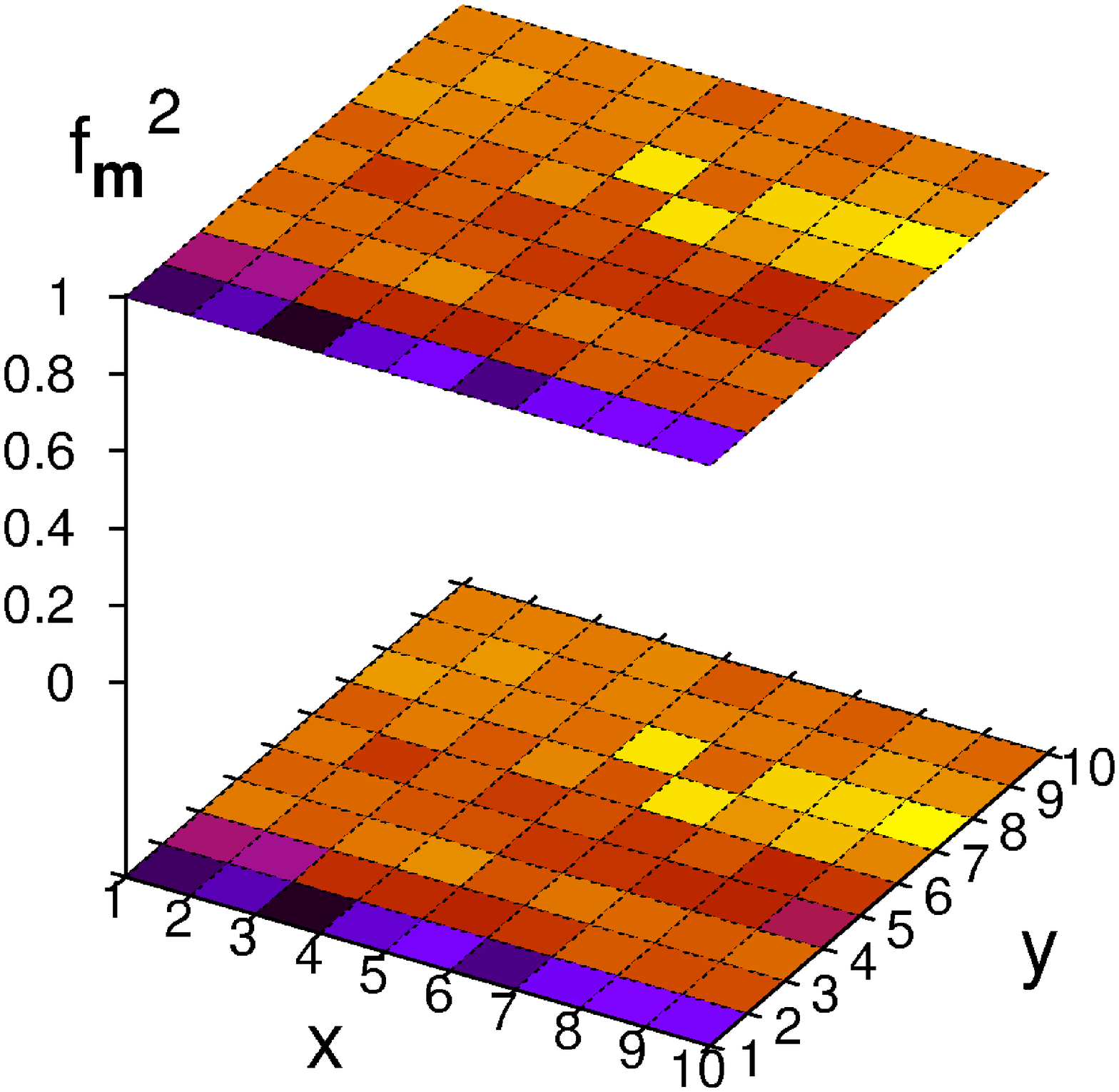,height=4.7cm}
\end{center}
\vspace{-.5cm}
\caption{\small Plots of a) the dispersion and the condensate wavefunctions and densities respectively for b) and c) $u_{00}=0.02$, d) and e) $u_{00}=0.02$ and $u_{01}=0.01$, f) and g) $u_{00}=0.02$ and $u_{02}=0.02$  in the unconventional case where $t_2=-0.6$ and $t_3=-0$.  Colours are assigned to the contour plots for b)-g) according to value the point at the foremost corner of a square. No scale is given for c) and g) since the changes in the size of the density are $\ll 1$ and due to numerical noise.}
\label{fig:Dcond2}
\end{figure}

For the unconventional case, further decreasing $t_2$ results in a growing potential wall between the minima of the C-phase (compare Figure \ref{fig:Denergy2} c) with Figure \ref{fig:Denergy3} a), for example).  As one would expect from the preceding argument, this entails that  $u^{in}_{00}$ must increase.

\section{Modulations for small potentials}\label{sec:results}

\begin{figure}
\begin{center}
a)
\epsfig{file=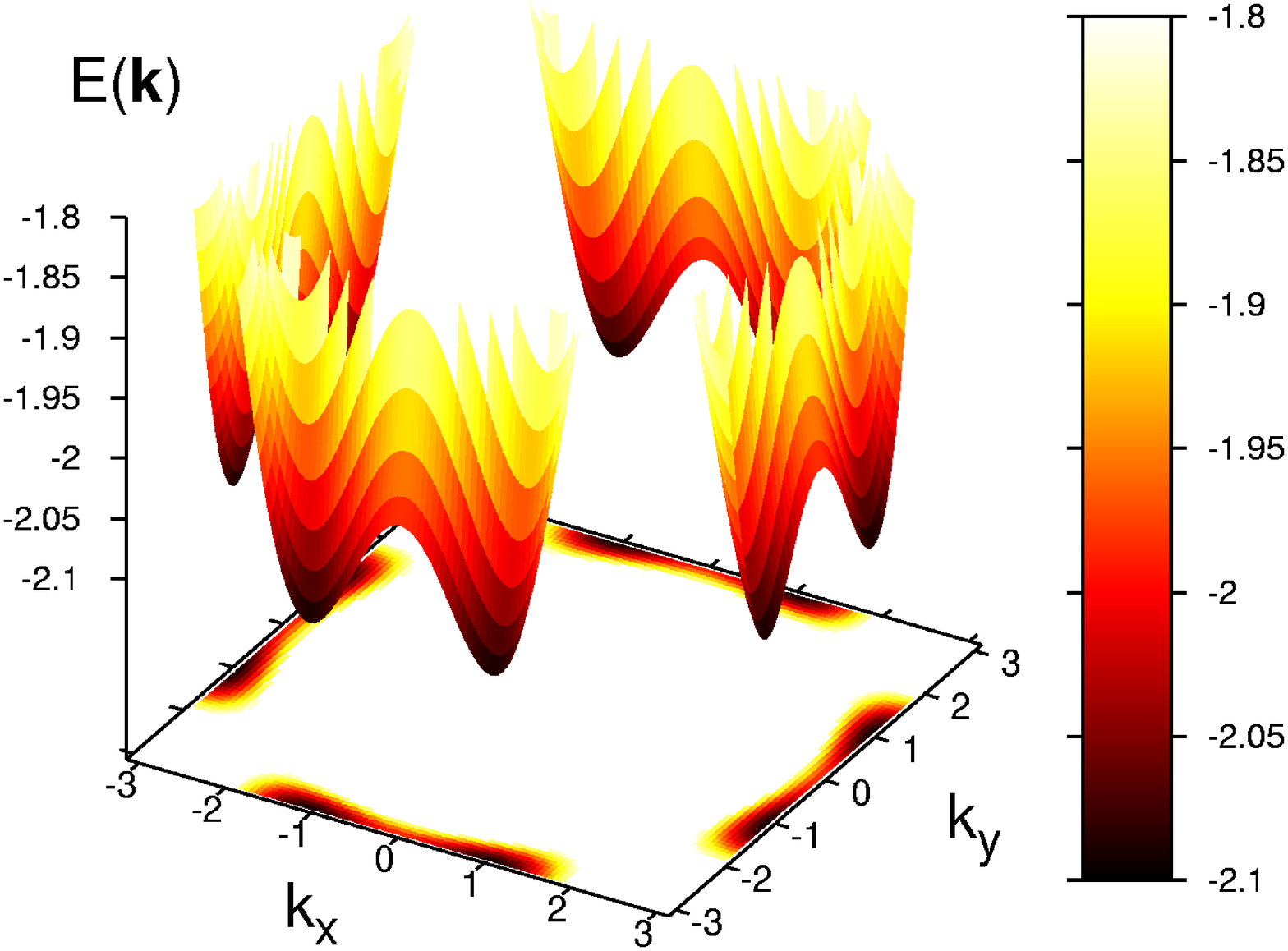,height=4.7cm}
\end{center}
\begin{center}
b)
\epsfig{file=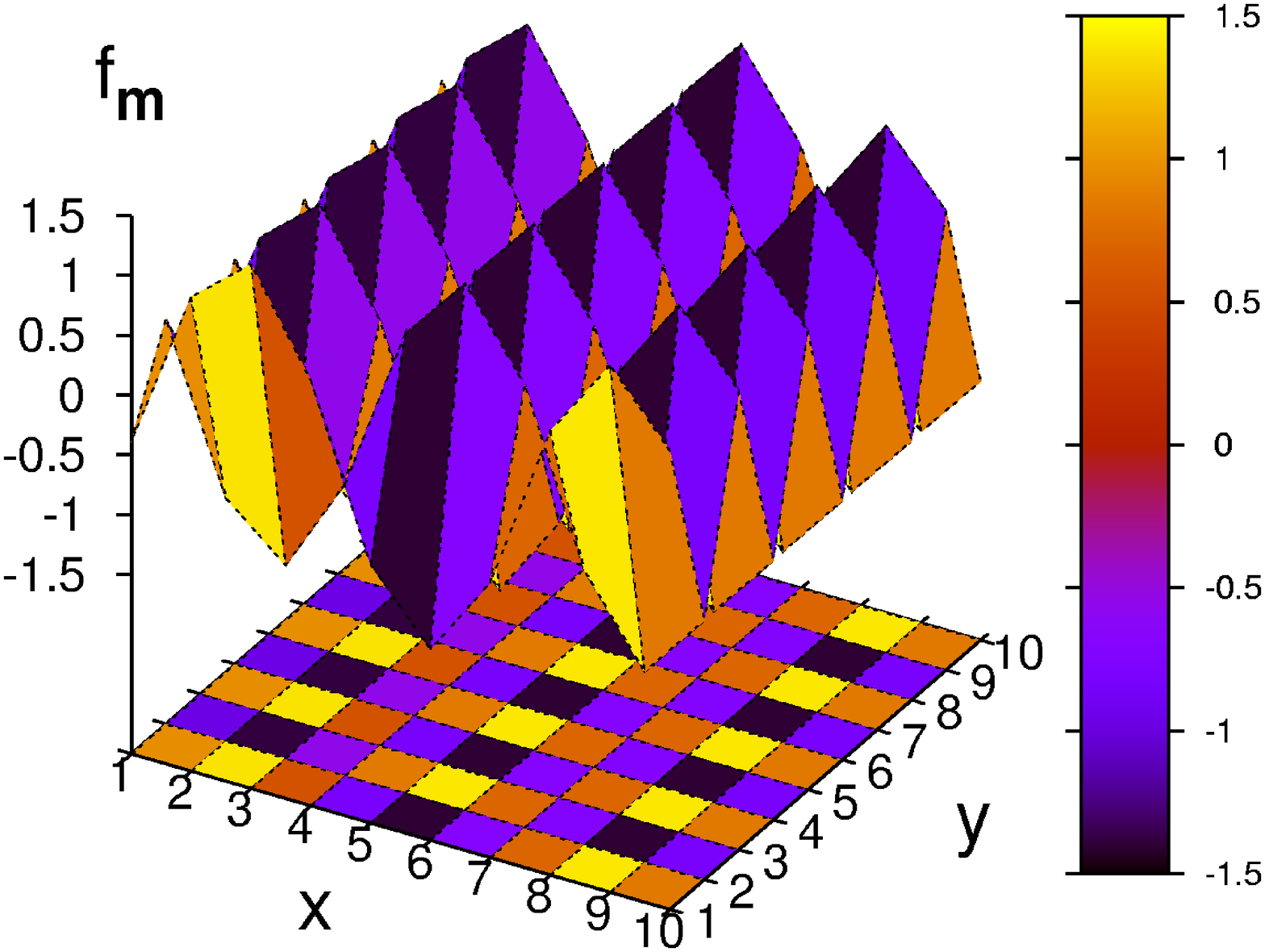,height=4.7cm}
\hspace{.3cm}
c)\epsfig{file=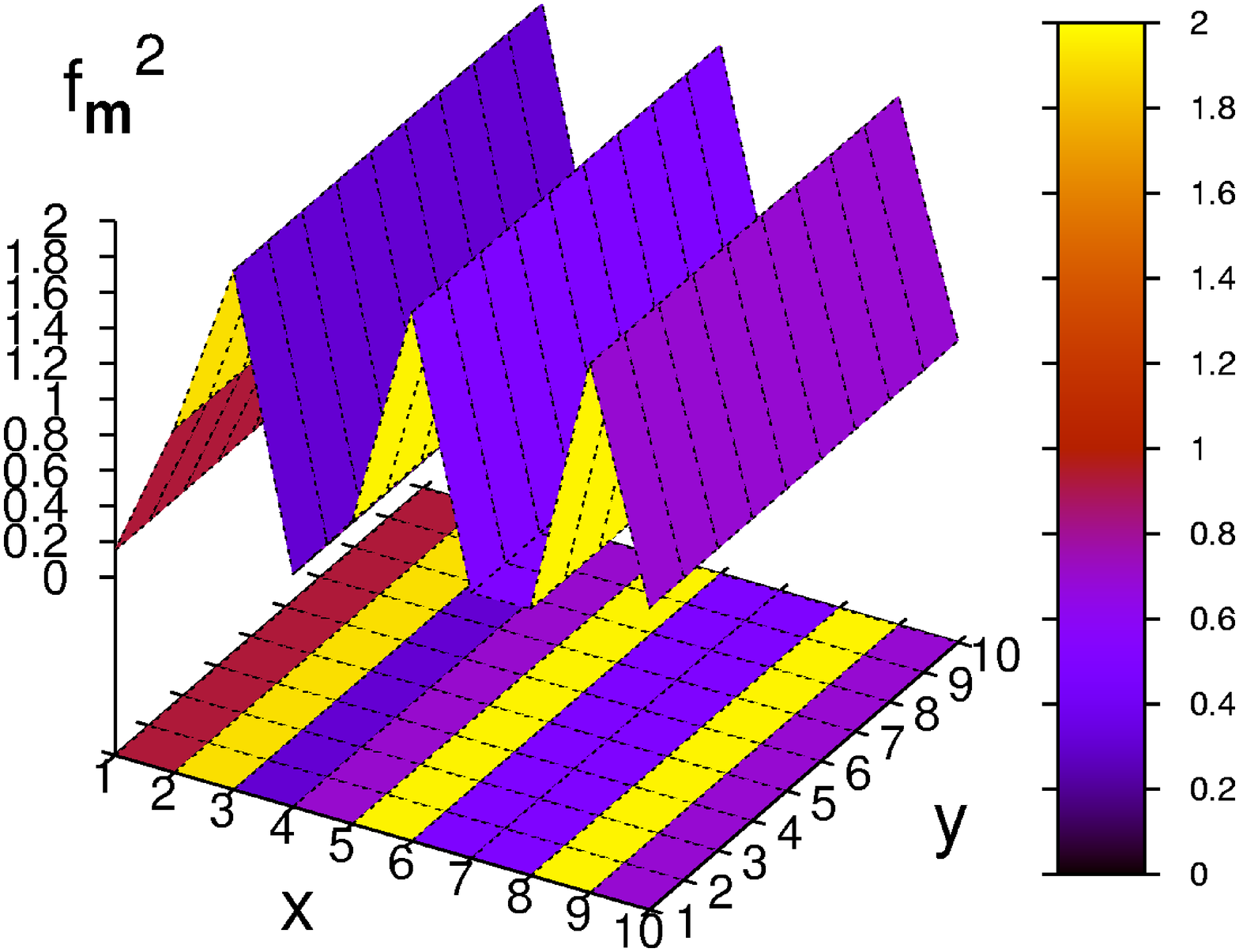,height=4.7cm}
\end{center}
\begin{center}
d)
\epsfig{file=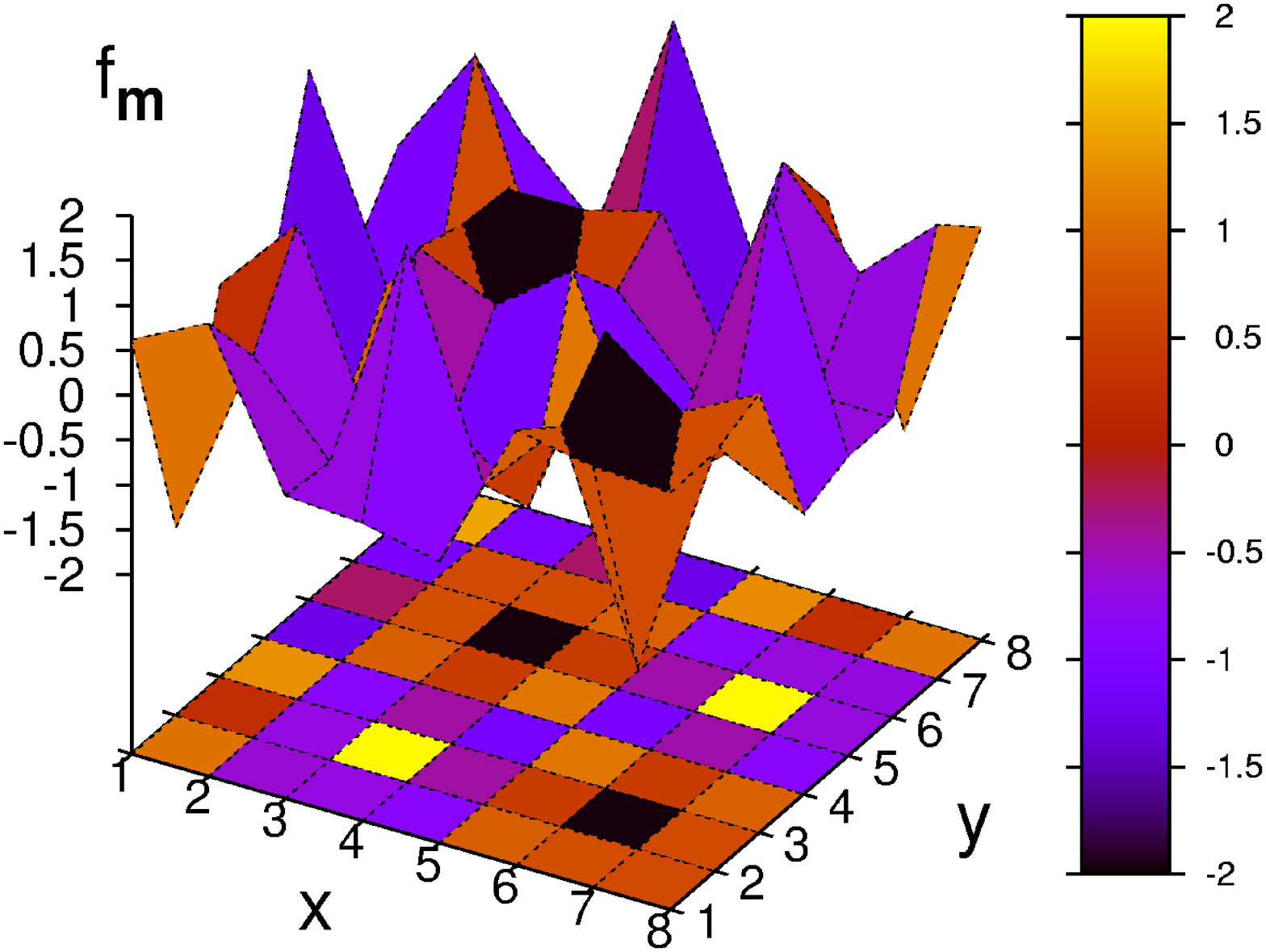,height=4.7cm}
\hspace{.3cm}
e)\epsfig{file=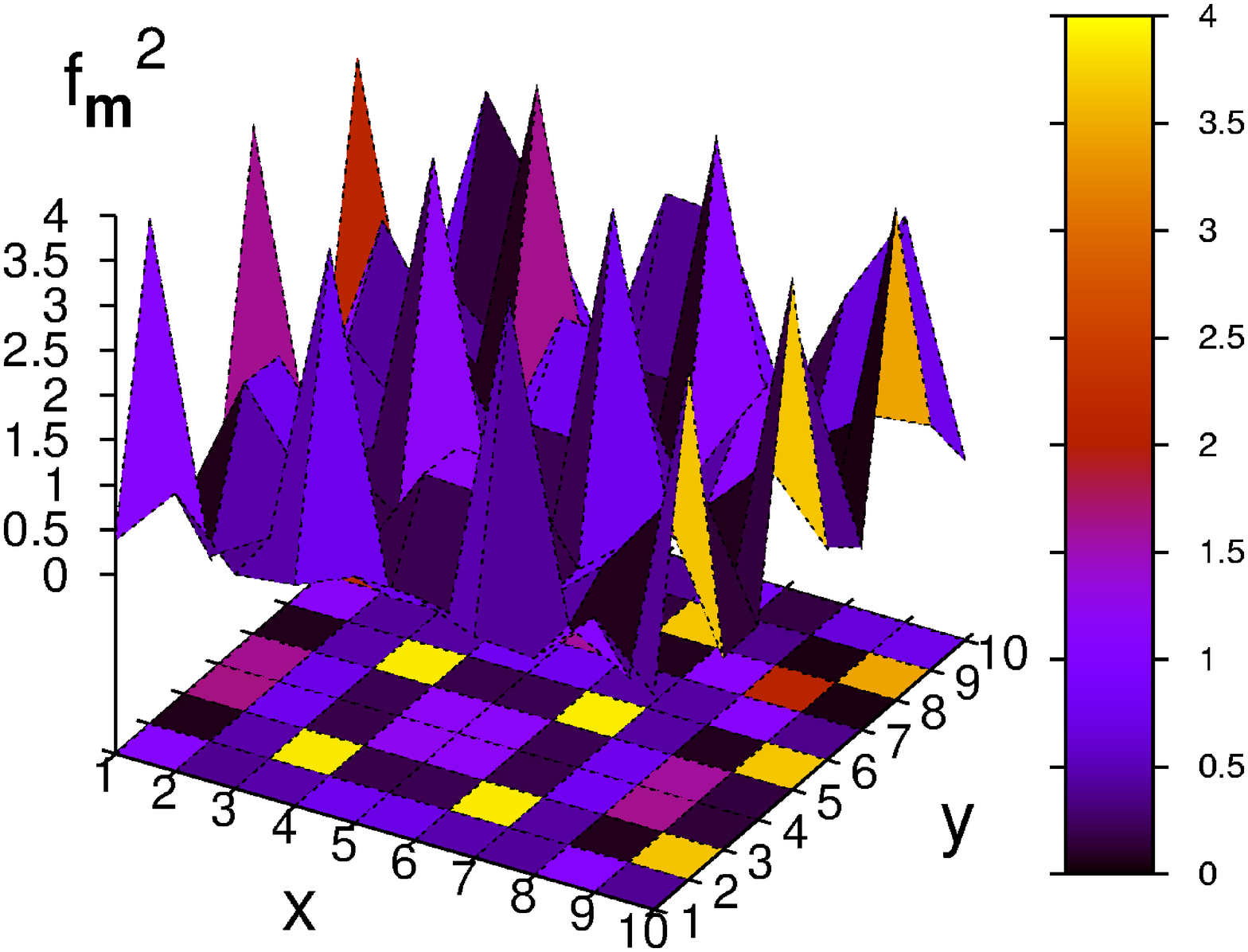,height=4.7cm}
\end{center}
\begin{center}
f)
\epsfig{file=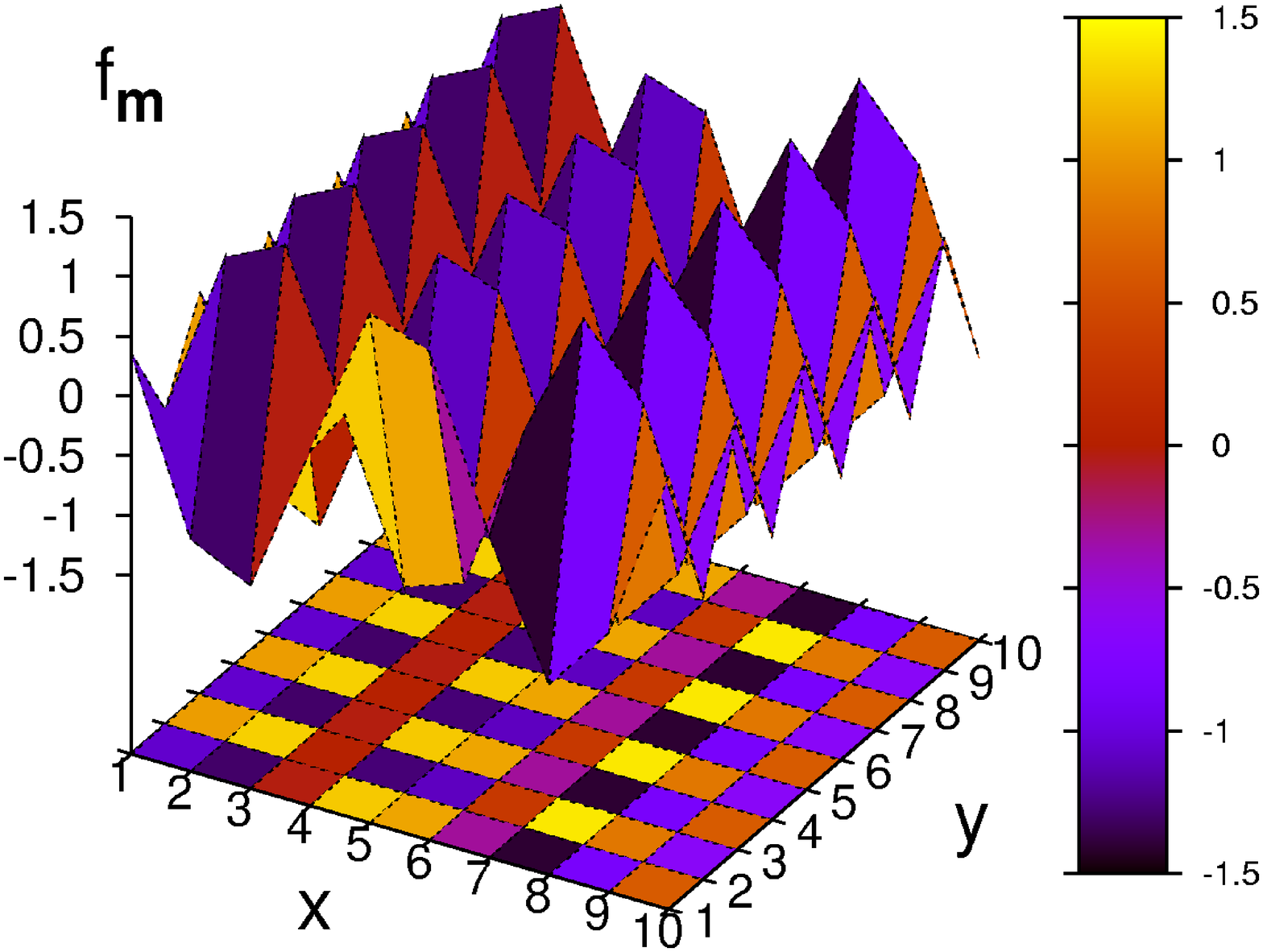,height=4.7cm}
\hspace{.3cm}
g)\epsfig{file=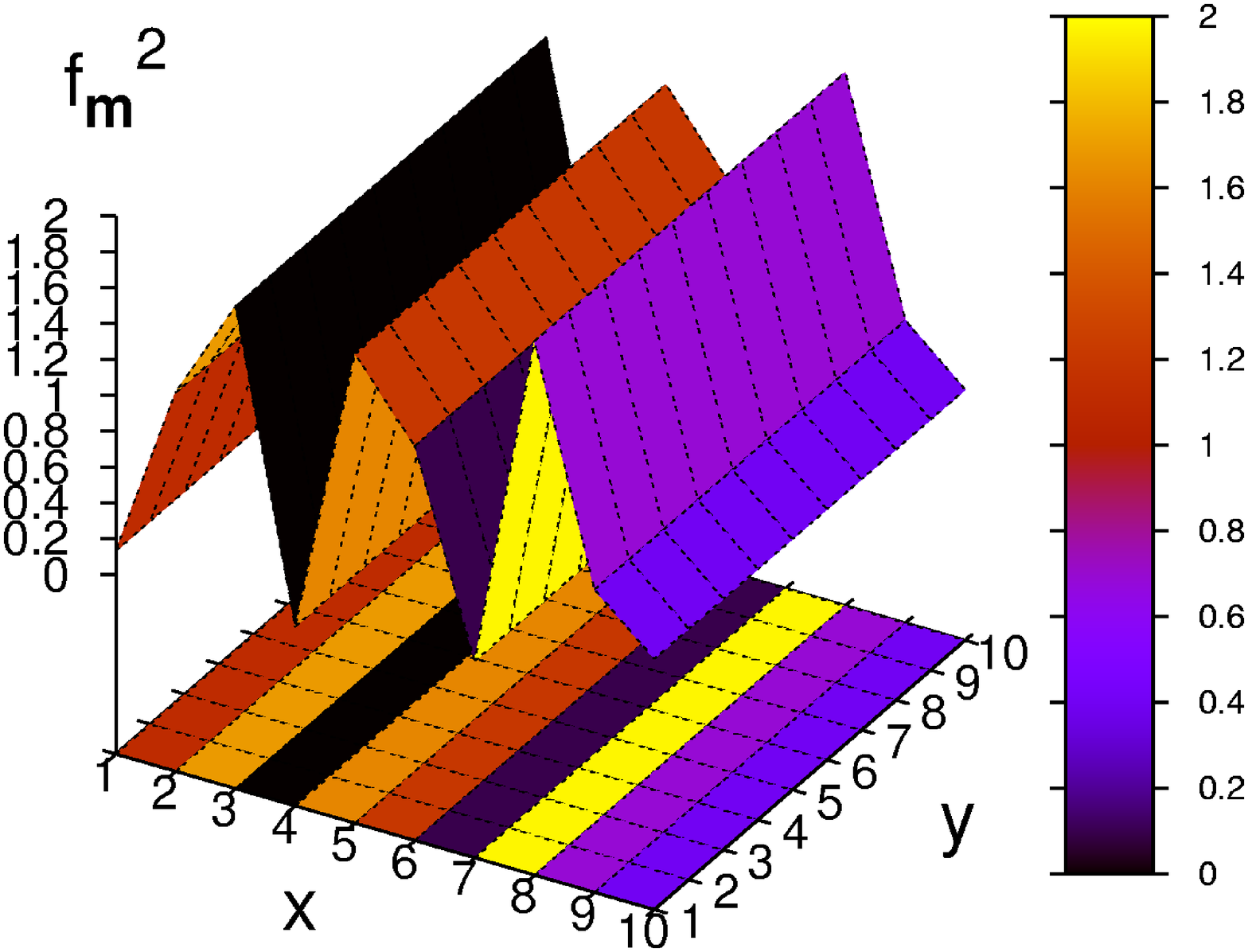,height=4.7cm}
\end{center}
\vspace{-.5cm}
\caption{\small  Plots of a) the dispersion and the condensate wavefunctions and densities respectively for b) and c) $u_{00}=0.02$, d) and e) $u_{00}=0.02$ and $u_{01}=0.01$, f) and g) $u_{00}=0.02$ and $u_{02}=0.02$  in the unconventional case where $t_2=-0.6$ and $t_3=-0.1$.  Colours are assigned to the contour plots for b)-g) according to value of the point at the foremost corner of a square.}
\label{fig:Dcond3}
\end{figure}

We have performed the numerical minimisation procedure of \S\ref{sec:procedure} for a number of values of the hopping parameters and small values of the repulsions.  Figures \ref{fig:Dcond1},  \ref{fig:Dcond2} and \ref{fig:Dcond3} shows results for parameter sets within the D, C and R phases respectively, where we have chosen a unconventional type dispersion since this is likely to be the most physically relevant.     

These results make the effects of the various forms of repulsion clear. In the non-homogeneous phases, a small repulsion causes the condensate wavefunction in Wannier space to take the form of a superposition respecting parity and time reversal symmetry, for which there are three possibilities, $f_{\bf m}\propto\cos({\bf k_1\cdot m})$, $f_{\bf m}\propto\cos({\bf k_2\cdot m})$ and $f_{\bf m}\propto\cos({\bf k_1\cdot m})\pm\cos({\bf k_2\cdot m})$.  The first two correspond to a `striped' condensate and the last corresponds to a `checkerboard' condensate. Our figures show that a finite value of the nearest-neighbour repulsion $u_{01}$ is needed for the selection of a checkerboard wavefunction in the C phases, whereas in the D phase a finite value of the next-nearest-neighbour repulsion  $u_{02}$ is required.  Otherwise, we have incommensurate striped wavefunctions in the D phase, and commensurate striped wavefunctions resulting in a homogeneous condensate density in the C phase. The R-phase follows the same selection criteria for the form of its wave-function as does the C-phase, however using the ${\bf k^\pm_{1,2}}$ instead.  In addition, the  wave-functions of both stripe and checkerboard phases are {\em staggered} due to the portions of the wave vectors that are located at the edges of the Brillouin zone; an conventional type dispersion would lack this staggering.  

It should be noted that the observed real-space modulations in our figures reflect the beating of the condensate period with that of the lattice.  

\begin{figure}[htb]
\begin{center}
a)
\epsfig{file=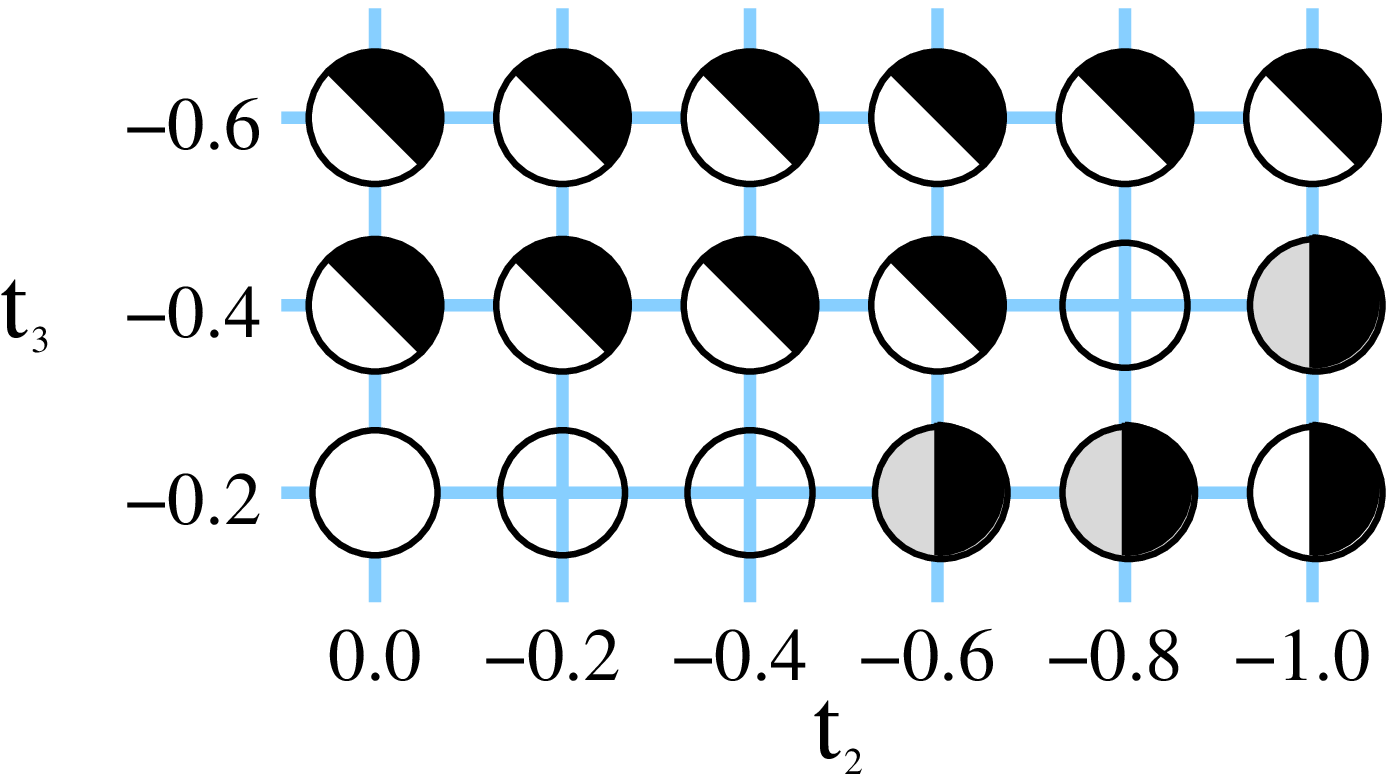,width=6.5cm}
\hspace{.3cm}
b)
\epsfig{file=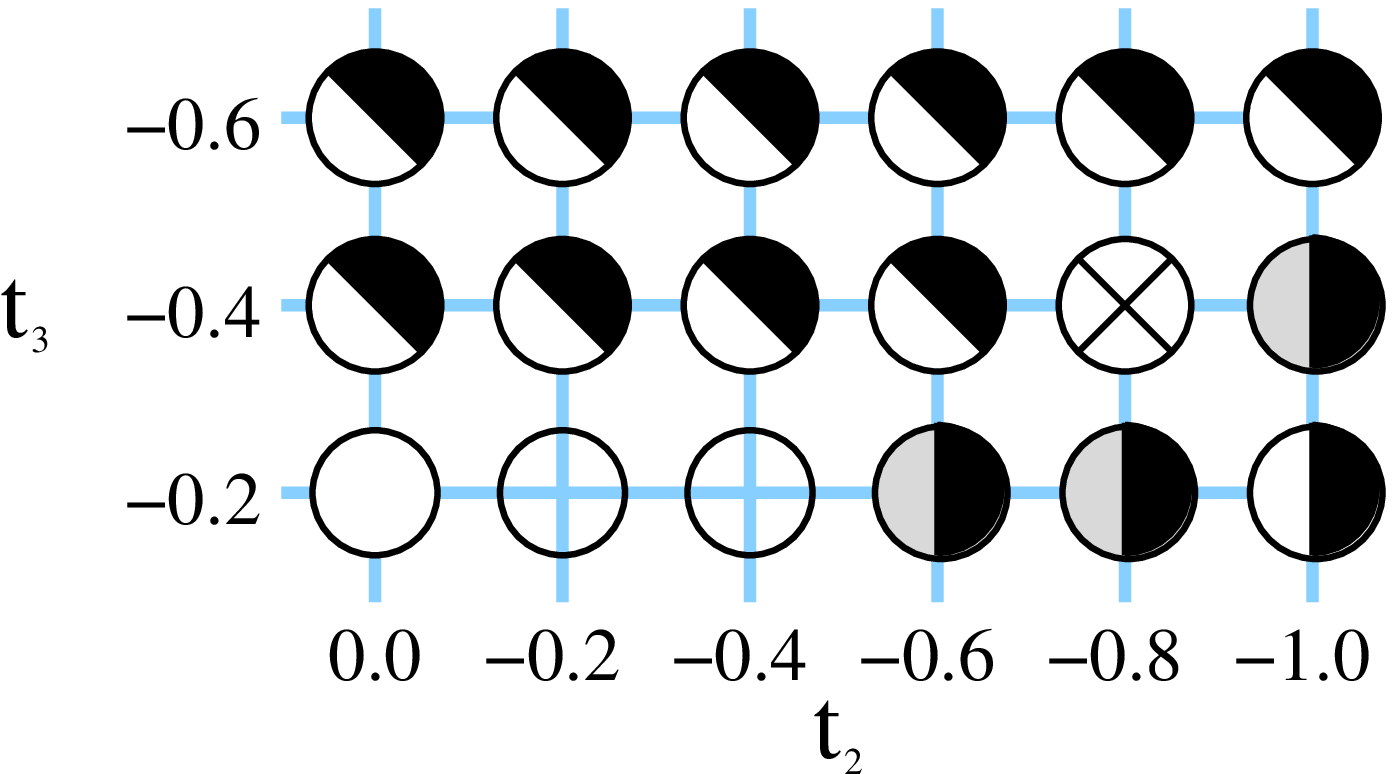,width=6.5cm}
\end{center}
\begin{center}
c)
\epsfig{file=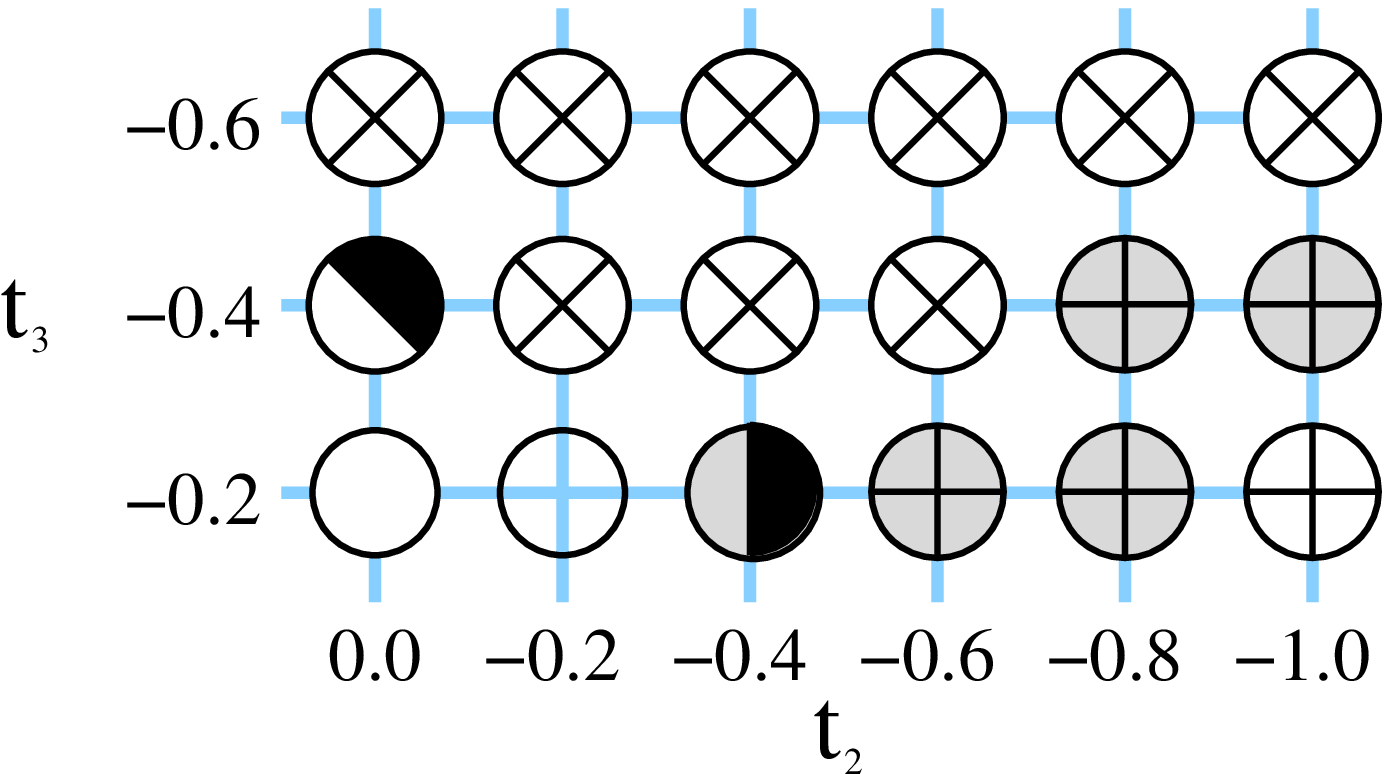,width=6.5cm}
\hspace{.3cm}
d)
\epsfig{file=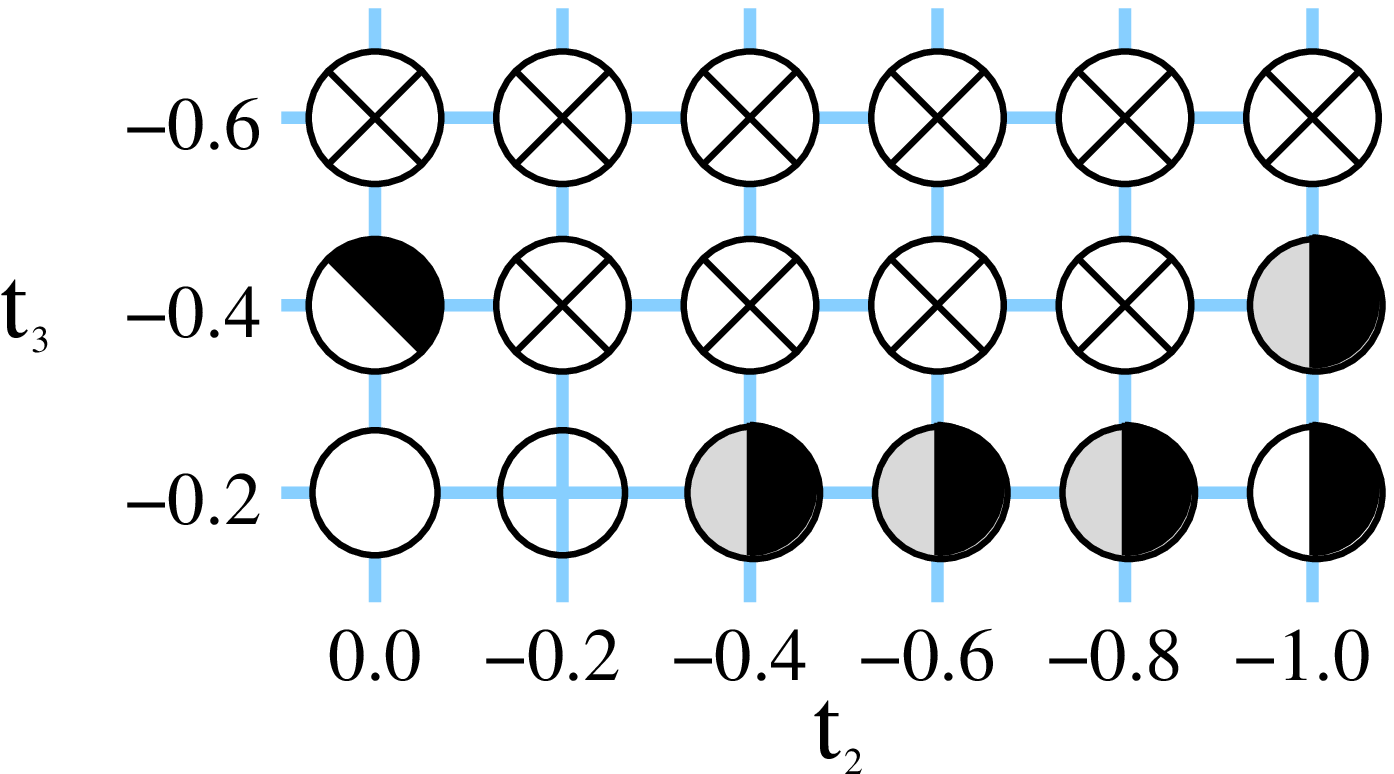,width=6.5cm}
\end{center}\begin{center}
e)
\epsfig{file=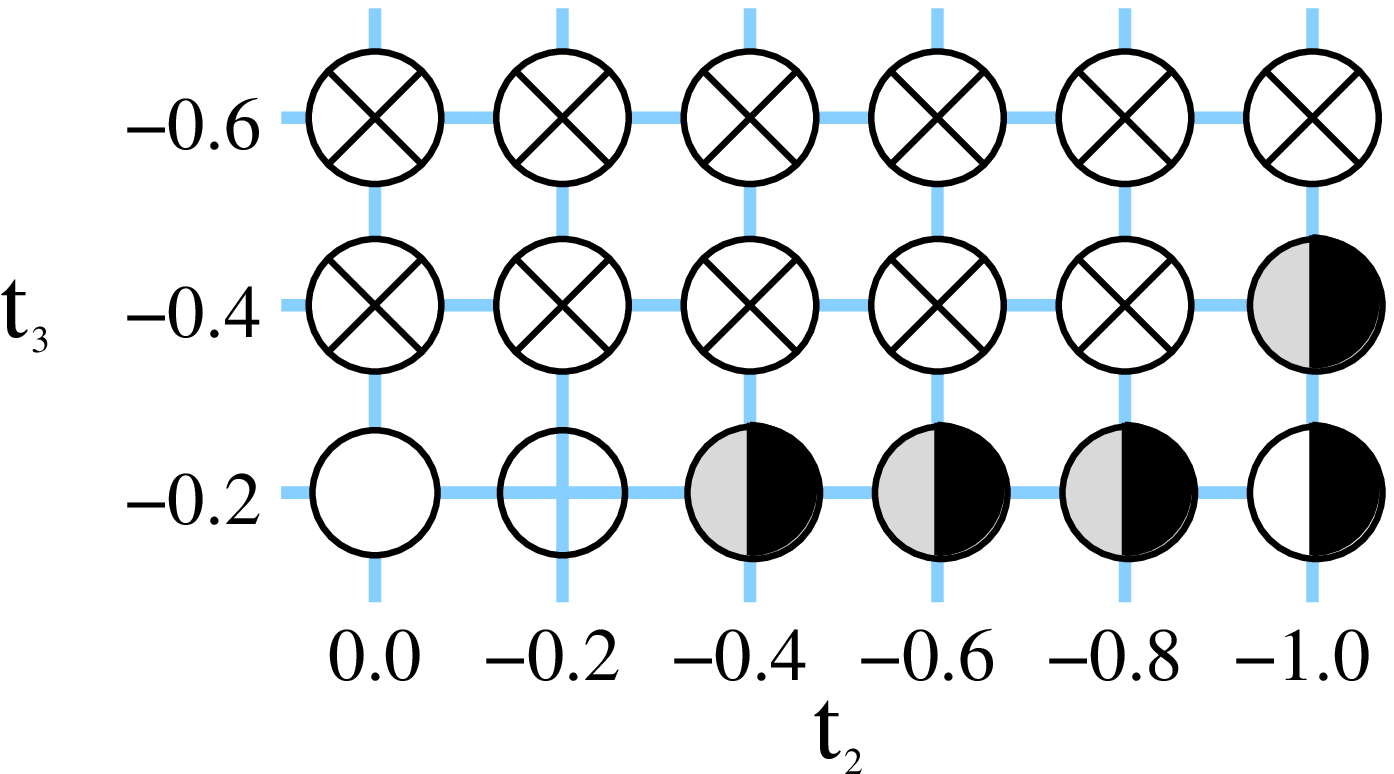,width=6.5cm}
\hspace{.3cm}
f)
\epsfig{file=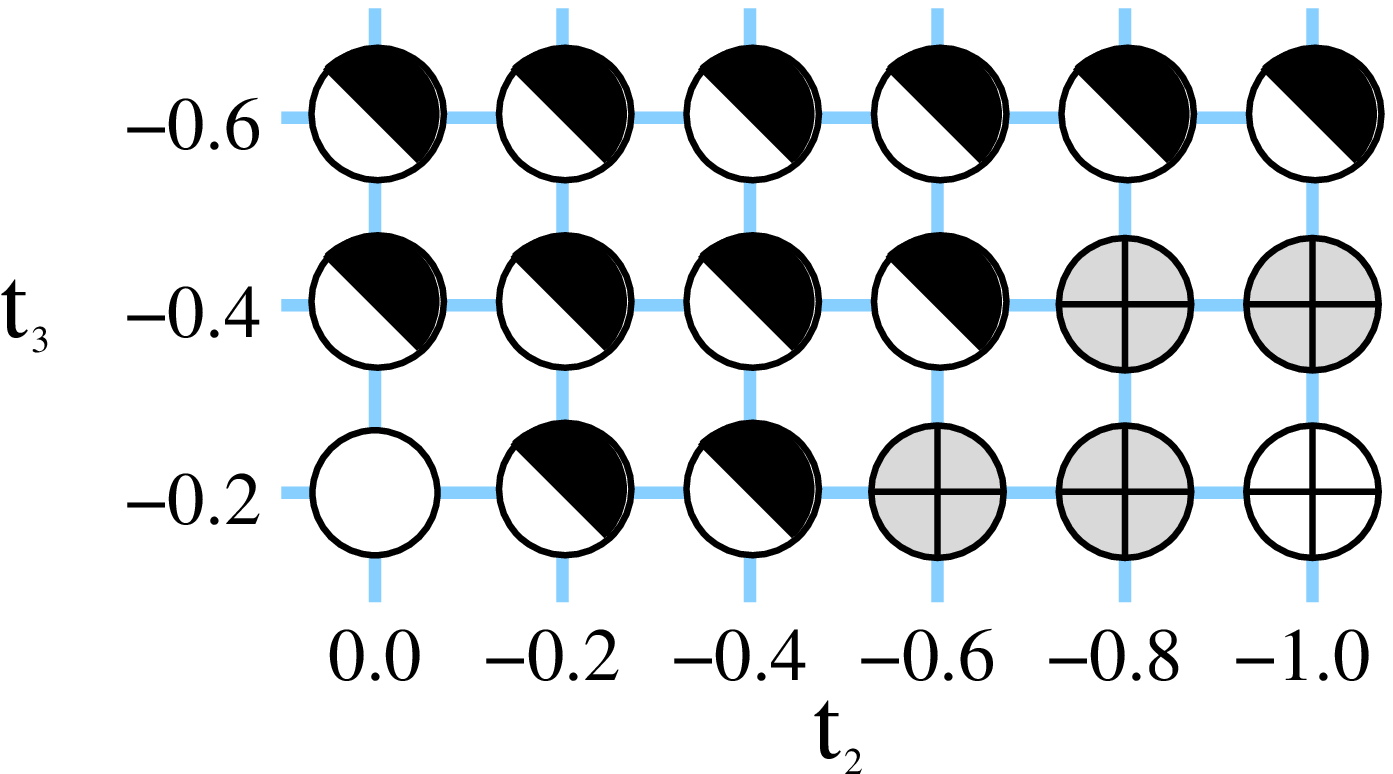, width=6.5cm}
\end{center}
\vspace{-.5cm}
\caption{\small Scans of a region of the $t_2-t_3$ plane with a) $u_{00}=0.2$, $u_{01}=0.03$, $u_{02}=0.02$, b)$u_{00}=0.2$, $u_{01}=0.01$, $u_{02}=0.02$, c)$u_{00}=0.01$, $u_{01}=0.03$, $u_{02}=0.02$, d) $u_{00}=0.01$, $u_{01}=0.01$, $u_{02}=0.02$, e) $u_{00}=0.2$, $u_{01}=0.0$, $u_{02}=0.02$, and f) $u_{00}=0.01$, $u_{01}=0.01$, $u_{02}=0.0$. Filled white circles indicate a homogeneous condensate wave function, empty circles indicate a `mixed' case, diagonal half-circles and crossed circles indicate diagonal stripes and checkerboards respectively, and vertical half-circles and crosses indicate stripes and checkerboards aligned with the axes of the lattice.  In the latter cases, shading indicates that the stripe or checkerboard is incommensurate with the lattice.}
\label{fig:Chkr1}
\end{figure}

However, for combinations of values of the various repulsive terms, the relationship is not as straightforward. Figure \ref{fig:Chkr1} shows scans of a region in the $t_2-t_3$ plane (carried out on $25\times25$ lattices with cross-checks on $50\times50$ lattices as discussed in \S\ref{sec:procedure}) for various values of the repulsive terms in order to obtain an exploratory overview of the behaviour of the system in the unconventional case.  The potentials for these scans were selected with the following in mind: for $t_2<-0.5$ and $t_3=0$, we typically seem to require $u_{00}>4u_{01}$ in order to move from a checkerboard state to the commensurate striped state.

As a general rule, it seems that in order for a diagonal checkerboard to emerge, $u_{02}>u_{00}$, and that for an aligned (R and C phase) checkerboard state to emerge we must have $u_{01}>u_{02}$, though this last condition does not seem to prevent the formation of a diagonal checkerboard in much (but not all) of the D phase.  The $0\ge  t_2\ge -0.2$, $t_3=-0.4$ region appears to remain striped in the presence of a finite $u_{01}$ potential regardless of the value of $u_{02}$.  We may also observe the presence of `mixed' ground states at certain values of $t_2$ and $t_3$; these correspond to the potential `wall' separating different minima being shallow.  This likely entails that in these areas the behaviour of the system is governed both by the potential terms and by the hopping terms in a complex fashion; this interplay would be worth further investigation.

The tendency of the $t_2=-0.8$, $t_3=-0.4$ region to exhibit behaviours characteristic of both D and R phases is likely due to its intermediate character, this behaviour is also be apparent in the $t_2=-0.4$, $t_3=-0.2$ region.  However, the latter (when not in a `mixed' phase) displays D-like behaviour when the former display R-like behaviour, and vice-versa.  It is possible that there are slopes or other differences in the valleys of the dispersion of the their that are too small to be observed on the scale of our plots of the dispersion that may account for this.

We are confident that our algorithm has in most cases located the appropriate ground states of the energy landscape (subject to the remarks in \S\ref{sec:procedure}), since they display behaviours consistent with what we would expect from Figure \ref{fig:Phase1}.  We should note, however, that in the vicinity of the transition to a mixed phase the minimum corresponding to an ordered phase could be very close to that corresponding to a mixed phase, and the algorithm may have some difficulty in locating the true minimum.  Therefore the results for $t_2=-0.2$, $t_3=-0.2$ should perhaps be treated with some caution.

 We might add that given the sensitivity of the form of the condensate wavefunction to that of the potential, it would be interesting to observe the effects of the inclusion of a long range interaction -- as in the case of a dipolar gas (such as in \cite{Santos:2000,O'Dell:2004,Glaum:2007} for example) or a magnetic field in a charged Bose liquid \cite{Alexandrov:1999,Alexandrov:2005}.  The latter would certainly be an interesting line of enquiry, given their role in a recently proposed explanation \cite{Alexandrov:2008a} of the  anomalous magneto-oscillations recently observed in cuprate superconductors \cite{Yelland:2008,Bangura:2008}.  In addition, simulations of the full quantum Bose Hubbard model with finite positive nearest-neighbour hopping \cite{Chen:2008} indicate possible quantum phase transitions to supersolid phases with finite values of the wave-vector.  It will certainly prove worthwhile to determine any connection between the transitions found here and those that might exist in the full Bose Hubbard model with negative values of the nearest-neighbour and next-to-nearest neighbour hopping parameters.

\section{Conclusion}

In the foregoing, we have calculated numerical solutions of the Gross-Pitaevskii equation in the tight-binding limit with additional hopping parameters and non-linear repulsive terms.  Building on the results of \cite{Alexandrov:2008}, we have described in more detail the mechanism of kinetic energy driven phase transitions in the superfluid phase, locating a new phase in the process.   We have also found that the value of $u_{00}$ at which one might expect an instability towards an inhomogeneous background to develop is altered by the variation of $t_2$.  This effect should be visible even in the case of conventional dispersions.  Finally, we have also presented results showing the effects of the repulsive terms on the modulation of the condensate in the various phases, as well as performing an exploratory mapping of the parameter space for various values of the repulsive terms.

We have discovered a new phase characterised by a wave vector which is incommensurate with the lattice in one direction and staggered in the other.  In addition, we have found that the value of $u_{00}$ at which the condensate becomes unstable is affected by form of the dispersion, and have confirmed that the form of the condensate wavefunction and density in the non-homogeneous phase is sensitive to the form of the repulsion. Work remains to be done regarding the effects of competing repulsive terms, in which analytic approaches may prove helpful.

\section{Acknowledgements}

The authors wish to thank J. F. Annett, J. Larson, V. V. Kabanov, J. Samson, and S. Savel'ev for valuable discussions.  This work was funded by EPSRC Grant No EP/D035589.


\end{document}